\documentclass{article}
\usepackage{spconf,amsmath,graphicx,xcolor,cleveref}		

\usepackage{enumitem}
\setlist{nosep, leftmargin=14pt}

\usepackage{mwe} 

\usepackage{bm}
\usepackage{amsfonts}
\usepackage{multirow}

\usepackage{subcaption}

\title{Nonlinear Equivariant Imaging: Learning Multi-Parametric Tissue Mapping without Ground Truth for Compressive Quantitative MRI\vspace{-.2cm}}
\name{\hspace{-0.21cm}Ketan Fatania, Kwai Y. Chau, Carolin M. Pirkl, Marion I. Menzel, Peter Hall and Mohammad Golbabaee \vspace{-2cm}\thanks{KF, KYC, PH and MG are with the Department of Computer Science at the University of Bath, UK: (KF432@bath.ac.uk). CMP and MIM are with GE Healthcare, Germany. MIM is also with the Department of Physics at the Technical University of Munich, Germany and AImotion Bavaria at the Technische Hochschule Ingolstadt, Germany.}}
\address{}
\begin{document}
\ninept
\maketitle
\begin{abstract}
Current state-of-the-art reconstruction for quantitative tissue maps from fast, compressive, Magnetic Resonance Fingerprinting (MRF), use supervised deep learning, with the drawback of requiring high-fidelity ground truth tissue map training data which is limited. This paper proposes NonLinear Equivariant Imaging (NLEI), a self-supervised learning approach to eliminate the need for ground truth for deep MRF image reconstruction. NLEI extends the recent Equivariant Imaging framework to nonlinear inverse problems such as MRF. Only fast, compressed-sampled MRF scans are used for training. NLEI learns tissue mapping using spatiotemporal priors: spatial priors are obtained from the invariance of MRF data to a group of geometric image transformations, while temporal priors are obtained from a nonlinear Bloch response model approximated by a pre-trained neural network. Tested retrospectively on two acquisition settings, we observe that NLEI (self-supervised learning) closely approaches the performance of supervised learning, despite not using ground truth during training.
\end{abstract}

\section{Introduction}
\label{sec:intro}

Magnetic Resonance Fingerprinting (MRF)~\cite{ref:ma2013}, is an accelerated Quantitative MRI (QMRI) method, for the acquisition of multi-parametric quantitative bio-property maps (QMaps) of the tissues,  in a single time-efficient scan. The reduced acquisition times are due to aggressive spatiotemporal subsampling, which leads to aliasing artefacts in the MRF image time-series data, and as a result QMaps. Current state-of-the-art for MRF image reconstruction use supervised deep learning e.g.~\cite{ref:fang2019oct, ref:golbabaee2021-lrtv_mrfresnet, ref:fatania2022}, for which training requires pairs of under-sampled, aliasing-contaminated MRF data, and their corresponding aliasing/artefact-free QMaps as ground truth. However, relying on ground truth is challenging, as: i) obtaining them requires long, clinically-infeasible scans, ii) long acquisitions are susceptible to motion artefacts (and correcting these may introduce interpolation artefacts), and iii) there is no real ground truth: each method for estimating ground truth QMaps comes with its own measurement effects and reconstruction artefacts, hence can only be considered a reference rather than real ground truth. 

Therefore, an alternative approach to supervised learning which does not rely on ground truth during training, would be highly beneficial. This work proposes a self-supervised deep learning approach, NonLinear Equivariant Imaging (NLEI), to enable MRF quantitative mapping (reconstruction of QMaps), using only fast compressive MRF scans as training data, without requiring ground truth QMaps. We also apply linear Equivariant Imaging (EI)~\cite{ref:chen2021}, the foundation for NLEI, for the first time to the MRF reconstruction problem. For competing algorithms, see~\cite{ref:gao2021, ref:chen2022}. NLEI learns a reconstruction mapping for the MRF \emph{nonlinear} inverse problem by incorporating spatiotemporal priors from the invariance of selected spatial transformations (e.g. rotations, flips) on estimated QMaps, and additionally {(unlike EI)} a differentiable model for the nonlinear Bloch response temporal dynamics approximated by a pre-trained neural network, BlochNet~\cite{ref:chen2020}. Tested retrospectively on two distinct MRF acquisitions, we observed that NLEI (self-supervised learning) closely approached the performance of supervised learning, despite not using ground truth during training.

\section{The MRF Inverse imaging Problem}
\label{sec:mrf}

MRF adopts a spatiotemporal compressed sensing acquisition:
\begin{equation}
	\label{eq:standard_linear_inverse_problem}
	\bm{y} \approx \bm{A}\bm{x}(\bm{q})
\end{equation}
where $ \bm{y} \in \mathbb{C}^{\,m\times T}$ are $m$ k-space measurements taken at $T$ timeframes, and $\bm{q}=\{\text{T1, T2, PD}\}$ are the unknown QMaps i.e. $n\times3$ images of the tissues' T1 and T2 relaxation times and Proton Density (PD) across $n>m$ voxels. The linear acquisition operator $ \bm{A}: \mathbb{C}^{\,n\times t}\rightarrow \mathbb{C}^{\,m\times T}$ models Fourier subsampling according to a set of temporally-varying k-space locations in each timeframe, combined with a temporal-domain SVD dimensionality reduction scheme~\cite{ref:mcgivney2014,ref:golbabaee2021-lrtv_mrfresnet} i.e., $3<t<T$. The Time-Series of Magnetisation Images (TSMI) for $n$ voxels and $t$ dimension-reduced timeframes are denoted by $ \bm{x}\in \mathbb{C}^{\,n\times t}$. The TSMIs' magnetisation responses (fingerprints) per voxel $v$, are \emph{nonlinearly} related to the tissue properties T1 and T2 relaxation times by the solutions of the Bloch differential equations, $\mathcal{B}$, scaled by the Proton Density, PD~\cite{ref:ma2013, jiang2015mr}:
\begin{equation}
	\label{eq:magnetisation_response}
	\bm{x}_{v} \approx \text{PD}_{v} \, \mathcal{B} \! \left( \text{T1}_{v}, \text{T2}_v \right)
\end{equation} 
The compressive nature of the acquisitions makes the estimation of QMaps ($\bm{q}$) from the undersampled MRF measurements ($\bm{y}$) a nonlinear ill-posed inverse problem~\eqref{eq:standard_linear_inverse_problem}.

The Bloch model can temporally constrain~\eqref{eq:standard_linear_inverse_problem}, but alone is absent of spatially-constraining priors to make the inverse problem well-posed. While supervised deep image reconstruction models can learn effective spatial priors from ground truth QMaps, i.e. interdependencies across image voxels, they would impose a significant scan-time challenge, as mentioned. We therefore build on the EI self-supervised learning framework to obtain a set of more generic (but still effective) spatially-constraining image priors, with the advantage of using only fast compressed MRF scans as training data.

%
\begin{figure*}[t]

	\begin{minipage}[b]{17.5cm}
  		\centering
  		\centerline{\includegraphics[scale=0.35]{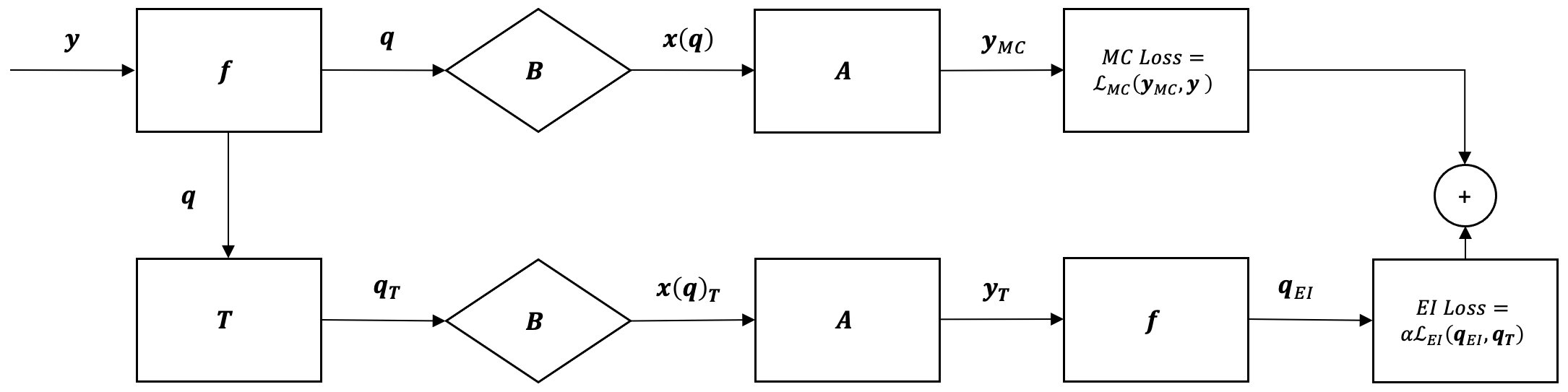}}
	\end{minipage}

	\caption{{Our NLEI algorithm for self-supervised learning of MRF quantitative mapping. See Non-Linear Equivariant Imaging for details.}}
	\label{fig:flowchart_nl_ei}

\end{figure*}

%

\section{Equivariant Imaging}
\label{sec:nl_ei}

Equivariant Imaging~\cite{ref:chen2021} exploits the assumption that an image (to be reconstructed) is invariant to certain types of transformations, e.g. reflections and rotations, in order to train a deep image reconstruction model ($\bm{f}$) in a self-supervised fashion. This is done by applying the acquisition operator $\bm{A}$, on the reconstructed transformed images obtained from $\bm{f}$, to yield new observations (k-space measurements) with information outside the range of the original observations. The new observations, once reconstructed by the same model $\bm{f}$, must result in a transformed image compared to the original reconstruction. 

While the original EI idea~\cite{ref:chen2021} works for linear inverse problems, i.e. estimating TSMIs ($\bm{x}$) and not QMaps ($\bm{q}$) in~\eqref{eq:standard_linear_inverse_problem}, it can be suboptimal for MRF by neglecting temporally-constraining  Bloch response priors that nonlinearly relate the TSMIs to lower-dimensional QMaps. Our NLEI algorithm builds on EI to additionally incorporate the Bloch priors~\eqref{eq:magnetisation_response} to estimate the QMaps in the nonlinear problem~\eqref{eq:standard_linear_inverse_problem}. Fig.\ref{fig:flowchart_nl_ei} shows the NLEI training pipeline.
\vspace{\baselineskip}
\\
\textbf{Non-Linear Equivariant Imaging:} 
The reconstruction model $\bm{\bm{f}(\bm{y})}:\bm{A}^H\bm{y}\rightarrow \bm{q}$ is a U-Net CNN following~\cite{ref:chen2021}, which learns a spatiotemporal mapping from aliased TSMIs, obtained from backprojected k-space measurements ($\bm{A}^H\bm{y}$), to the artefact-free QMaps. $\bm{A}^H$ is the adjoint of $\bm{A}$, and our experiments used the temporal dimension $t=10$ for the complex-valued TSMIs, leading to 20ch stacked real and imaginary parts for the inputs of $\bm{f}$, whereas outputs are 4ch QMaps of T1, T2, $\text{PD}_{\text{real}}$ and $\text{PD}_{\text{imag}}$. Training uses a weighted sum of two MSE losses $\mathcal{L}_{MC}+\alpha \mathcal{L}_{EI}$ ($\alpha>0$).

\begin{itemize}
\renewcommand\labelitemi{--}

\item \textbf{Measurement Consistency (MC) loss,} $\mathcal{L}_{MC}(\bm{y}_{MC}, \bm{y}) $\textbf{:} is a routinely-used loss in compressed sensing literature, first applied to deep MRF in~\cite{ref:chen2020} for minimising discrepancies between scanner k-space measurements $\bm{y}$, and those obtained from the reconstructed QMaps $\bm{q}:=\bm{f}(\bm{y})$, following the nonlinear forward model (1). To be specific, a pre-trained BlochNet model ($\bm{B}$) which approximates (2) was used to map $\bm{q}$ to TSMIs $\bm{x}(\bm{q})\approx \bm{B}(\bm{q})$ with $\text{PD}_{\text{real}}$ and $\text{PD}_{\text{imag}}$ used to obtain complex-valued TSMIs, followed then by the compressed subsampling operator $\bm{A}$, to obtain k-space data $\bm{y}_{MC}:= \bm{A}\circ\bm{B}(\bm{q})$. Minimising $\mathcal{L}_{MC}$ enables $\bm{f}$ to find an inverse mapping for $\bm{A}\circ\bm{B}$, where the reconstructed QMaps respect the forward model physics.

\item \textbf{EI loss} $\mathcal{L}_{EI}(\bm{q}_{EI}, \bm{q}_{\bm{T}}) $\textbf{:} minimises discrepancies between reconstructed $\bm{q}$, and the spatially-transformed reconstructed QMaps $\bm{q}_{EI}$, (Fig.\ref{fig:flowchart_nl_ei}). To be specific, spatial transformations $\bm{T}$, are applied to the QMaps  $\bm{q}:= \bm{f}(\bm{y})$, reconstructed from the original (scanner) k-space data. An approximate $\bm{A}\circ \bm{B}$ of the nonlinear forward model~\eqref{eq:standard_linear_inverse_problem} are applied to $\bm{q}_{\bm{T}}:= \bm{T}(\bm{q})$, to obtain new k-space measurements, which were then reconstructed by $\bm{f}$ into $\bm{q}_{EI}:=\bm f\circ \bm{A} \circ \bm{B} \circ \bm{T}(\bm{q})$. The EI loss enables $\bm{f}$ to learn a reconstruction mapping in a self-supervised manner that respects the image invariance properties, i.e. $\bm{f}$ should learn that $\bm{q}$ and $\bm{q}_{EI}$ are only different by a transformation: $\bm{T}(\bm{q}) \approx \bm{q}_{EI}$. \\
\end{itemize}
\textbf{Linear Equivariant Imaging:} The linear EI algorithm~\cite{ref:chen2021} for MRF can be reduced from the NLEI pipeline in Fig.\ref{fig:flowchart_nl_ei}: (i) let $\bm{\bm{f}(\bm{y})}:\bm{A}^H\bm{y}\rightarrow \bm{x}$ reconstruct a TSMI, and (ii) remove the BlochNet $\bm{B}$ (diamond-shapes in Fig.\ref{fig:flowchart_nl_ei}) responsible for the forward model's nonlinearity. The result is the EI algorithm to reconstruct an artefact-reduced TSMI $\bm{x}$, albeit uninformed/unconstrained by the Bloch response priors. An MRF dictionary-matching step~\cite{ref:ma2013} then can be used to estimate QMaps from the EI-reconstructed TSMI.
\vspace{\baselineskip}
\\
\textbf{BlochNet,} $\bm{B}$~\cite{ref:chen2020}\textbf{:} approximates (2) by a differentiable neural network model, that is kept frozen and used within the NLEI's training pipeline (Fig.\ref{fig:flowchart_nl_ei}) to add temporal Bloch response priors. Implemented by a CNN of 2 hidden layers (each with 300 filters, ReLU activations), BlochNet uses $1\times1$ filters to process QMaps (input) in a voxel-wise manner and output the corresponding TSMI i.e. $\bm{x}(q)\approx \bm{B}(\bm{q})$. This network is trained offline from NLEI. Training data uses an SVD dimension-reduced ($t=10$) FISP-MRF dictionary~\cite{ref:mcgivney2014} with 94,777 fingerprints, that are simulated Bloch responses (using EPG simulator~\cite{weigel2015extended}) for combinations of T1/T2 values in a logarithmically-sampled grid $($T1, T2$)\in[0.01, 6]\times [0.004, 4]$ (sec).%
\vspace{\baselineskip}
\\
\textbf{Transformations, $\bm{T}$:} An important component of EI is the selection of appropriate transformations to learn image invariances. For compressed sensing MRI, randomly selected rotations have been successfully used by EI~\cite{ref:chen2022}, while shift transformations have no benefit for Fourier based acquisitions~\cite{ref:chen2021}. For our work, the NLEI and EI use a random selection from 7 transformations defined by combinations of $90^{\circ}$ rotations and vertical flips, accounting for all orientations: 1) Vertical Flip, 2) $90^{\circ}$ Rotation, 3) Vertical Flip with $90^{\circ}$ Rotation, 4) $180^{\circ}$ Rotation, 5) Vertical Flip with $180^{\circ}$ Rotation, 6) $270^{\circ}$ Rotation and 7) Vertical Flip with $270^{\circ}$ Rotation. We limit rotation angles to multiples of $90^{\circ}$ to prevent interpolation artefacts.

\section{Numerical Experiments}
\label{sec:experiments}
%
\begin{figure}[t]

\begin{minipage}[b]{0.6125cm}
  \centering
  \centerline{\includegraphics[scale=0.4]{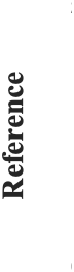}}
  \vspace{0.38cm}
\end{minipage}
\hfill
\begin{minipage}[b]{-3.05cm} 				
  \centering
  \centerline{\includegraphics[scale=0.4]{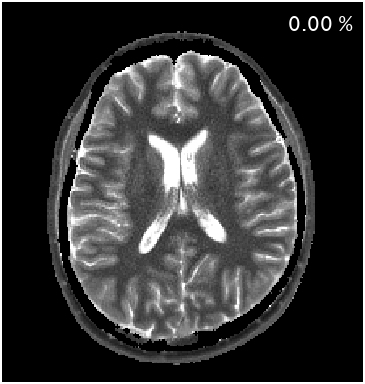}}
\end{minipage}
\hfill
\begin{minipage}[b]{-3.175cm}				
  \centering
  \centerline{\includegraphics[scale=0.4]{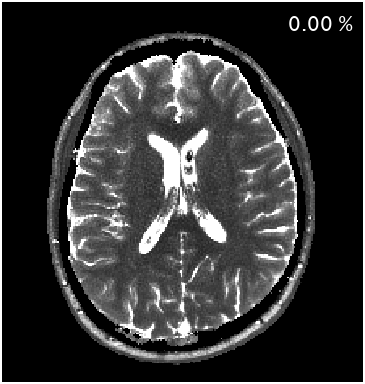}}
\end{minipage}
\hfill
\begin{minipage}[b]{-3.3cm}
  \centering
  \centerline{\includegraphics[scale=0.4]{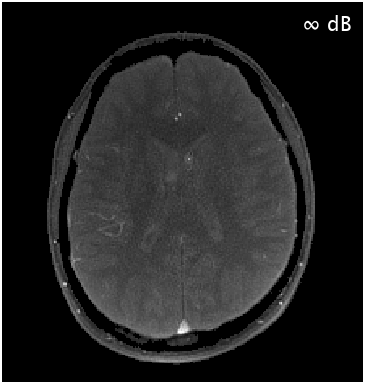}}
\end{minipage}
\\ 
\begin{minipage}[b]{0.6125cm}
  \centering
  \centerline{\includegraphics[scale=0.4]{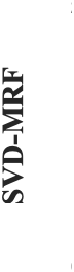}}
  \vspace{0.38cm}
\end{minipage}
\hfill
\begin{minipage}[b]{-3.05cm} 				
  \centering
  \centerline{\includegraphics[scale=0.4]{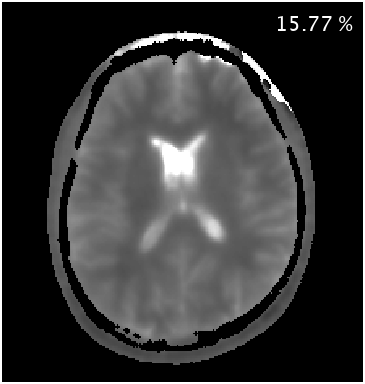}}
\end{minipage}
\hfill
\begin{minipage}[b]{-3.175cm}				
  \centering
  \centerline{\includegraphics[scale=0.4]{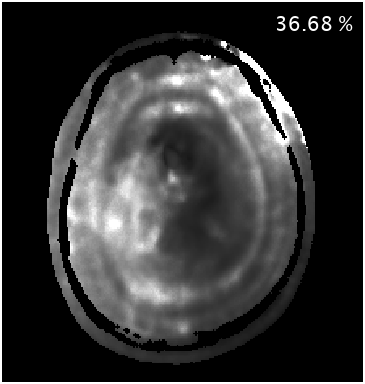}}
\end{minipage}
\hfill
\begin{minipage}[b]{-3.3cm}
  \centering
  \centerline{\includegraphics[scale=0.4]{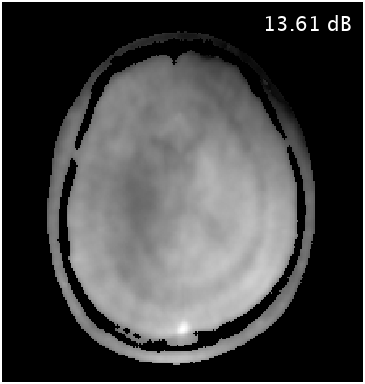}}
\end{minipage}
\\ 
\begin{minipage}[b]{0.6125cm}
  \centering
  \centerline{\includegraphics[scale=0.4]{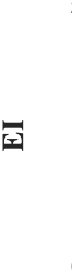}}
  \vspace{0.38cm}
\end{minipage}
\hfill
\begin{minipage}[b]{-3.05cm} 				
  \centering
  \centerline{\includegraphics[scale=0.4]{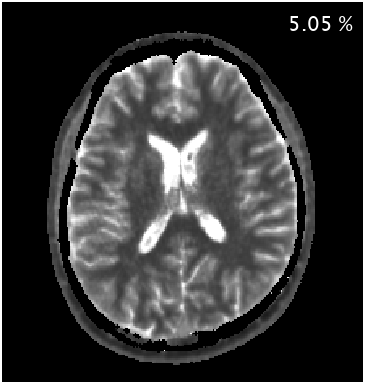}}
\end{minipage}
\hfill
\begin{minipage}[b]{-3.175cm}				
  \centering
  \centerline{\includegraphics[scale=0.4]{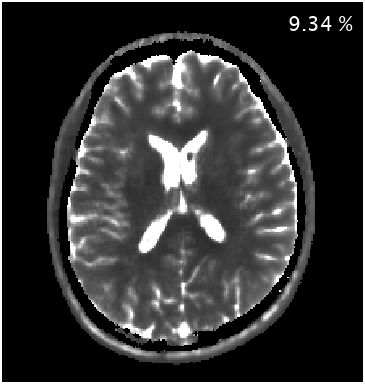}}
\end{minipage}
\hfill
\begin{minipage}[b]{-3.3cm}
  \centering
  \centerline{\includegraphics[scale=0.4]{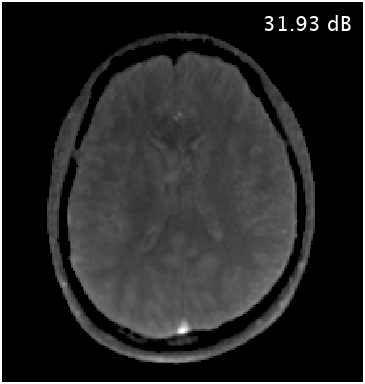}}
\end{minipage}
\\ 
\begin{minipage}[b]{0.6125cm}
  \centering
  \centerline{\includegraphics[scale=0.4]{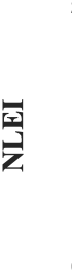}}
  \vspace{0.38cm}
\end{minipage}
\hfill
\begin{minipage}[b]{-3.05cm} 				
  \centering
  \centerline{\includegraphics[scale=0.4]{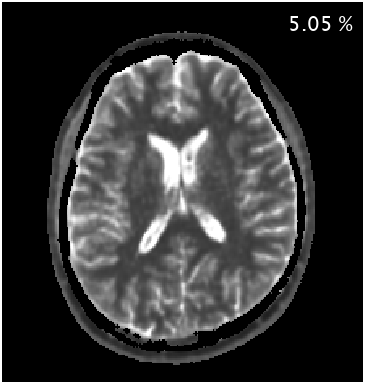}}
\end{minipage}
\hfill
\begin{minipage}[b]{-3.175cm}				
  \centering
  \centerline{\includegraphics[scale=0.4]{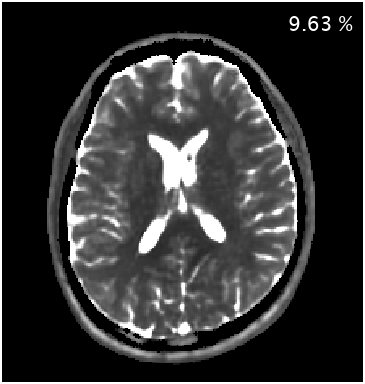}}
\end{minipage}
\hfill
\begin{minipage}[b]{-3.3cm}
  \centering
  \centerline{\includegraphics[scale=0.4]{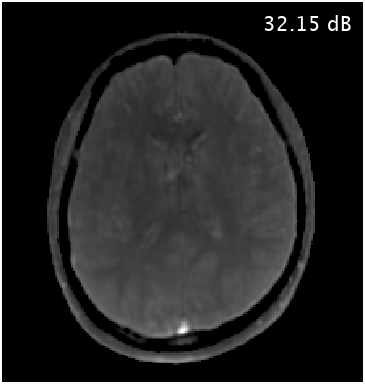}}
\end{minipage}
\\ 
\begin{minipage}[b]{0.6125cm}
  \centering
  \centerline{\includegraphics[scale=0.4]{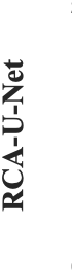}}
  \vspace{0.38cm}
\end{minipage}
\hfill
\begin{minipage}[b]{-3.05cm} 				
  \centering
  \centerline{\includegraphics[scale=0.4]{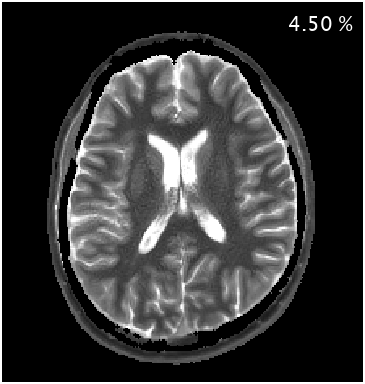}}
\end{minipage}
\hfill
\begin{minipage}[b]{-3.175cm}				
  \centering
  \centerline{\includegraphics[scale=0.4]{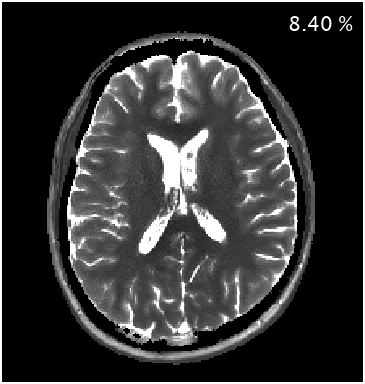}}
\end{minipage}
\hfill
\begin{minipage}[b]{-3.3cm}
  \centering
  \centerline{\includegraphics[scale=0.4]{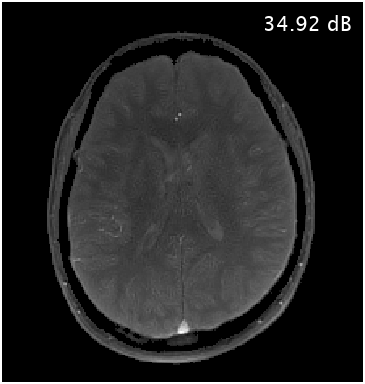}}
\end{minipage}
\\ 
\begin{minipage}[b]{3.5cm}
  \centering
  \vspace{1mm}
  \centerline{\includegraphics[scale=0.4]{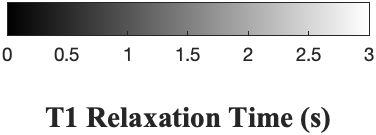}}
\end{minipage}
\hfill
\begin{minipage}[b]{-3.35cm}
  \centering
  \centerline{\includegraphics[scale=0.4]{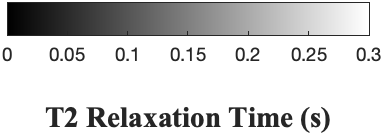}}
\end{minipage}
\hfill
\begin{minipage}[b]{-3.325cm}
  \centering
  \centerline{\includegraphics[scale=0.4]{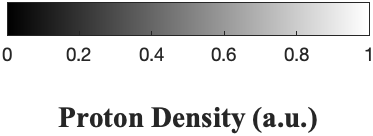}}
\end{minipage}
\caption{Tissue Map Results using Spiral Subsampling for Slice 10 of 15, with MAPE (\%) for T1 and T2, and PSNR (dB) for PD.}
\label{fig:results_spiral}
\vspace{-\baselineskip}
%
\end{figure}

\begin{figure}[t]

\begin{minipage}[b]{0.6125cm}
  \centering
  \centerline{\includegraphics[scale=0.4]{./figures/vertical_west_labels/reference_t1_colorbar_vertical_west_cropped.png}}
  \vspace{0.38cm}
\end{minipage}
\hfill
\begin{minipage}[b]{-3.05cm}
  \centering
  \centerline{\includegraphics[scale=0.4]{./figures/qmaps/ground_truth/gt_t1_with_mask_caxis0to3.png}}
\end{minipage}
\hfill
\begin{minipage}[b]{-3.175cm}
  \centering
  \centerline{\includegraphics[scale=0.4]{./figures/qmaps/ground_truth/gt_t2_with_mask_caxis0to0.3.png}}
\end{minipage}
\hfill
\begin{minipage}[b]{-3.3cm}
  \centering
  \centerline{\includegraphics[scale=0.4]{./figures/qmaps/ground_truth/gt_pd_with_mask_caxis0to1.png}}
\end{minipage}
\\ 
\begin{minipage}[b]{0.6125cm}
  \centering
  \centerline{\includegraphics[scale=0.4]{./figures/vertical_west_labels/svd-mrf_t1_colorbar_vertical_west_cropped.png}}
  \vspace{0.38cm}
\end{minipage}
\hfill
\begin{minipage}[b]{-3.05cm}
  \centering
  \centerline{\includegraphics[scale=0.4]{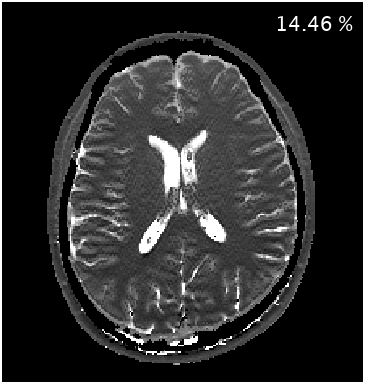}}
\end{minipage}
\hfill
\begin{minipage}[b]{-3.175cm}
  \centering
  \centerline{\includegraphics[scale=0.4]{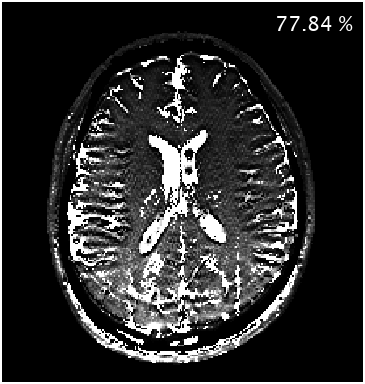}}
\end{minipage}
\hfill
\begin{minipage}[b]{-3.3cm}
  \centering
  \centerline{\includegraphics[scale=0.4]{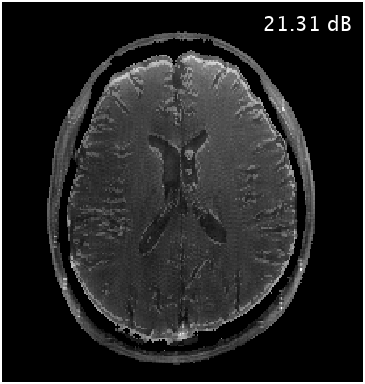}}
\end{minipage}
\\ 
\begin{minipage}[b]{0.6125cm}
  \centering
  \centerline{\includegraphics[scale=0.4]{./figures/vertical_west_labels/ei_t1_colorbar_vertical_west_cropped.png}}
  \vspace{0.38cm}
\end{minipage}
\hfill
\begin{minipage}[b]{-3.05cm}
  \centering
  \centerline{\includegraphics[scale=0.4]{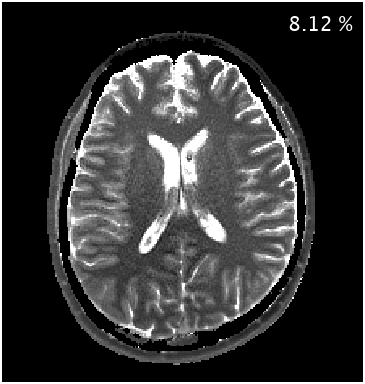}}
\end{minipage}
\hfill
\begin{minipage}[b]{-3.175cm}
  \centering
  \centerline{\includegraphics[scale=0.4]{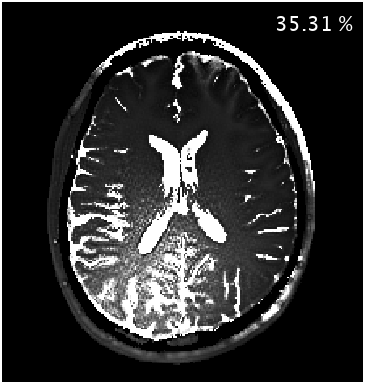}}
\end{minipage}
\hfill
\begin{minipage}[b]{-3.3cm}
  \centering
  \centerline{\includegraphics[scale=0.4]{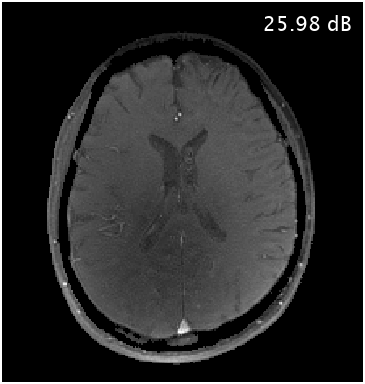}}
\end{minipage}
\\ 
\begin{minipage}[b]{0.6125cm}
  \centering
  \centerline{\includegraphics[scale=0.4]{./figures/vertical_west_labels/nlei_t1_colorbar_vertical_west_cropped.png}}
  \vspace{0.38cm}
\end{minipage}
\hfill
\begin{minipage}[b]{-3.05cm}
  \centering
  \centerline{\includegraphics[scale=0.4]{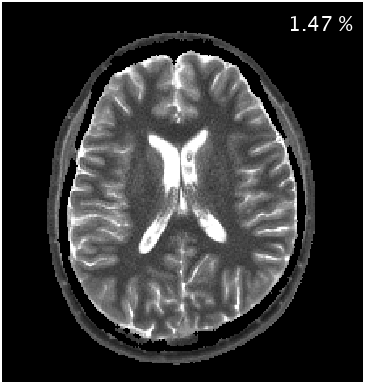}}
\end{minipage}
\hfill
\begin{minipage}[b]{-3.175cm}
  \centering
  \centerline{\includegraphics[scale=0.4]{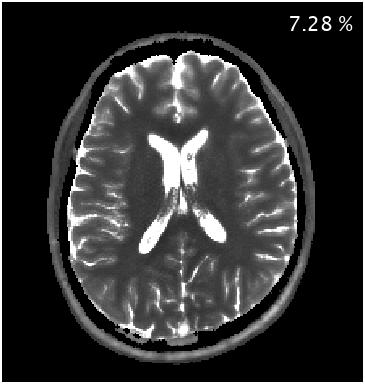}}
\end{minipage}
\hfill
\begin{minipage}[b]{-3.3cm}
  \centering
  \centerline{\includegraphics[scale=0.4]{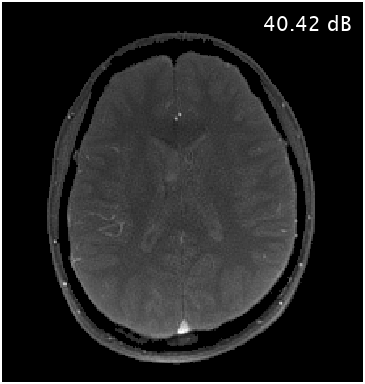}}
\end{minipage}
\\ 
\begin{minipage}[b]{0.6125cm}
  \centering
  \centerline{\includegraphics[scale=0.4]{./figures/vertical_west_labels/rca-u-net_t1_colorbar_vertical_west_cropped.png}}
  \vspace{0.38cm}
\end{minipage}
\hfill
\begin{minipage}[b]{-3.05cm}
  \centering
  \centerline{\includegraphics[scale=0.4]{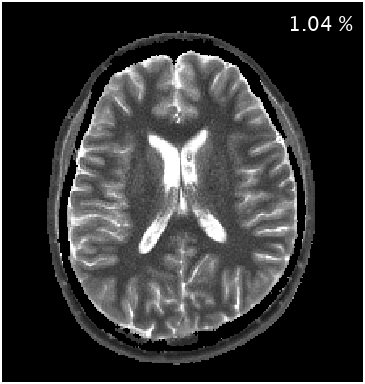}}
\end{minipage}
\hfill
\begin{minipage}[b]{-3.175cm}
  \centering
  \centerline{\includegraphics[scale=0.4]{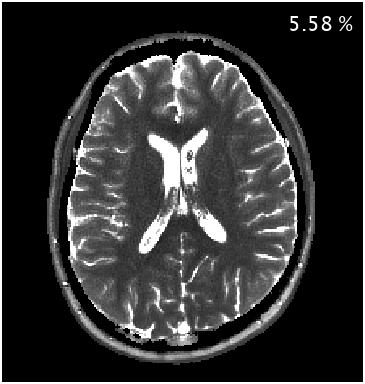}}
\end{minipage}
\hfill
\begin{minipage}[b]{-3.3cm}
  \centering
  \centerline{\includegraphics[scale=0.4]{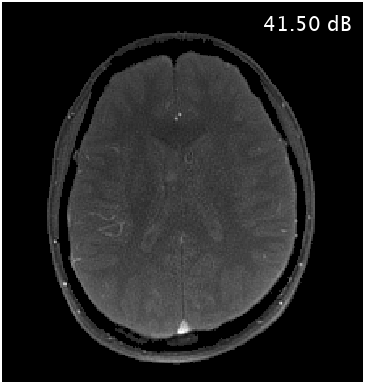}}
\end{minipage}
\\ 
\begin{minipage}[b]{3.5cm}
  \centering
  \vspace{1mm}
  \centerline{\includegraphics[scale=0.4]{./figures/qmaps/ground_truth/gt_t1_with_mask_caxis0to3_colorbar_horizontal_south.png}}
\end{minipage}
\hfill
\begin{minipage}[b]{-3.35cm}
  \centering
  \centerline{\includegraphics[scale=0.4]{./figures/qmaps/ground_truth/gt_t2_with_mask_caxis0to0.3_colorbar_horizontal_south.png}}
\end{minipage}
\hfill
\begin{minipage}[b]{-3.325cm}
  \centering
  \centerline{\includegraphics[scale=0.4]{./figures/qmaps/ground_truth/gt_pd_with_mask_caxis0to1_colorbar_horizontal_south.png}}
\end{minipage}
\caption{Tissue Map Results using EPI Subsampling for Slice 10 of 15, with MAPE (\%) for T1 and T2, and PSNR (dB) for PD.}
\label{fig:results_epi}
\vspace{-\baselineskip}
%
\end{figure}

%
\textbf{Dataset:} We used a dataset of T1, T2 and PD QMaps of 2D axial brain scans from 8 subjects across 15 slices. Complex-valued TSMIs and k-space MRF data were retrospectively simulated from these QMaps via~\eqref{eq:magnetisation_response} and~\eqref{eq:standard_linear_inverse_problem}, correspondingly. The reference QMaps and TSMIs had spatial dimensions of $n=224 \times 224 $ pixels with head-masks applied, which were generated using the Proton Density. For the Bloch response model a truncated FISP-MRF protocol~\cite{jiang2015mr} was used with $T=200$ repetitions i.e. 5 times less (accelerated) than the original FISP-MRF. For MRF k-space data, we simulated single-coil acquisitions using two distinct cartesian (FFT) k-space subsampling patterns evolving across each temporal frame: (i) a rotating Spiral, as in~\cite{ref:chen2020} and (ii) shifting horizontal lines using multi-shot Echo Planar Imaging (EPI)~\cite{benjamin2019multi}. We sampled $m=771$ k-space locations in each timeframe, corresponding to a spatial compression ratio of 65:1. The TSMIs were dimension-reduced ($t=10$) following~\cite{ref:mcgivney2014}. The dataset was split into 105 slices from 7 subjects for training, and 15 slices from the $\text{8}^{\text{th}}$ subject for testing. 
\vspace{.2cm}
\\
\textbf{Tested Algorithms:} We compared the performance of the proposed NLEI to EI~\cite{ref:chen2021}, SVD-MRF~\cite{ref:mcgivney2014} and RCA-U-Net~\cite{ref:fang2019oct} baselines. SVD-MRF is a non-data driven approach which reconstructs backprojected (aliased) TSMI $\bm{A}^H\bm{y}$ from k-space data, followed by MRF dictionary-matching to estimate QMaps. Other tested algorithms use deep learning to further process the backprojected TSMIs to produce aliasing-reduced estimations. RCA-U-Net is a state-of-the-art deep supervised learning MRF model that uses ground truth QMaps for training. Separate RCA-U-Net models were trained for T1, T2, $\text{PD}_{\text{real}}$ and $\text{PD}_{\text{imag}}$ following~\cite{ref:fang2019oct} using 1000 epochs, L1 loss, and linearly decaying learning rate from 200 to 1000 epochs. On the other hand, NLEI and EI are self-supervised learning models that do not use ground truth for training (they only use under-sampled MRF k-space data). NLEI and EI used the U-Net as in~\cite{ref:chen2021}, without the residual connection between the initial and final layers, and trained using 1000 epochs, batch size 2, Adam optimiser, weight decay $10^{-8}$, initial learning rate $5\times10^{-4}$ decreasing by factor 10 at 300 epochs. We used 3 randomly selected transformations per iteration from the 7 previously defined, and applied them across each batch to create new batch sizes of 6. Optimal values for $\alpha$ were found experimentally: $10^{-8}$ for NLEI Spiral, $10^{-4}$ for NLEI EPI, $10^{-5}$ for EI Spiral, and $10^{-2}$ for EI EPI.
\vspace{.2cm}
\\
\textbf{Evaluation Metrics:} We used the Mean Absolute Error (MAE), Mean Absolute Percentage Error (MAPE), Peak Signal-to-Noise-Ratio (PSNR) and Structural Similarity Index Measure (SSIM). Head-masks were applied to all reconstructions and metrics were then calculated and averaged across 15 test slices.
\vspace{.2cm}
%
\begin{table*}[t] 
	
	\centering
	
	\resizebox{16cm}{!}{	
	
	\begin{tabular}{ c c || c c c c || c c c c } 
		
		\hline \hline \multicolumn{2}{ c ||}{} & \multicolumn{4}{ c ||}{Spiral} & \multicolumn{4}{ c }{EPI} \\ \hline \hline
		
		 & & SVD-MRF & EI & NLEI & RCA-U-Net & SVD-MRF & EI & NLEI & RCA-U-Net \\ \hline \hline
		 
		\multirow{ 2}{*}{MAE (s)}
		& T1 & 0.1371 & 0.0577 & 0.0532 & 0.0473 & 0.1658 & 0.1096 & 0.0180 & 0.0121 \\
		& T2 & 0.0484 & 0.0178 & 0.0162 & 0.0134 & 0.1004 & 0.0451 & 0.0139 & 0.0094 \\ 
		\hline
		
		\multirow{ 2}{*}{MAPE (\%)} 
		& T1 & 12.2519 & 5.0527 & 4.2236 & 3.8465 & 13.5611 & 7.4711 & 1.2951 & 0.9502 \\
		& T2 & 34.9255 & 9.3400 & 8.6272 & 7.3329 & 68.3580 & 33.5549 & 6.7534 & 5.2095 \\
		\hline		
		
		\multirow{ 4}{*}{PSNR (dB)} 
		& TSMI & 11.3135 & 26.0396 & - & - & 17.1590 & 3.2324 & - & - \\
		& T1 & 23.9384 & 33.2161 & 33.6557 & 34.7350 & 22.4085 & 23.3885 & 39.2725 & 42.9782 \\
		& T2 & 25.7089 & 34.8714 & 36.7658 & 38.3916 & 20.6887 & 31.1333 & 37.0679 & 41.1515 \\
		& PD & 13.9547 & 31.9296 & 31.4679 & 34.0273 & 21.5492 & 25.9406 & 38.2944 & 39.7831 \\ 
		\hline
		
		\multirow{ 4}{*}{SSIM} 
		& TSMI & 0.5979 & 0.7721 & - & - & 0.5819 & 0.5542 & - & - \\
		& T1 & 0.8195 & 0.9409 & 0.9425 & 0.9558 & 0.8962 & 0.9230 & 0.9898 & 0.9967 \\
		& T2 & 0.8469 & 0.9514 & 0.9537 & 0.9688 & 0.7529 & 0.8376 & 0.9642 & 0.9838 \\
		& PD & 0.7676 & 0.8896 & 0.8913 & 0.9224 & 0.7659 & 0.9050 & 0.9837 & 0.9906 \\ 
		\hline \hline
	
	\end{tabular}
	
	}
	
	\caption[...]{Evaluation metrics, averaged over 15 test slices, show increasing performance for each acquisition scheme, moving from SVD-MRF~\cite{ref:mcgivney2014}, EI~\cite{ref:chen2021}, NLEI (ours) to RCA-U-Net~\cite{ref:fang2019oct}. NLEI (self-supervised learning) performs closest to RCA-U-Net (supervised learning).}
	\label{tab:results_table}


\end{table*}
%
\\
\textbf{Results and Discussion:} Metrics in Table~\ref{tab:results_table}, for both subsampling patterns, show a clear progression in performance from SVD-MRF to EI, NLEI and RCA-U-Net, with NLEI (self-supervised approach trained without ground truth) approaching the performance of the supervised model RCA-U-Net. The results in Fig.\ref{fig:results_spiral} show similar performance for NLEI and EI, while Fig.\ref{fig:results_epi} shows NLEI outperforms SVD-MRF and EI, while being close to RCA-U-Net. SVD-MRF, NLEI and EI exhibit blurring, which can also be seen in RCA-U-Net to a lesser extent. This is intrinsic to spiral subsampling due to prioritised sampling of low frequencies at the centre of k-space. 

We acknowledge that no noise was used in our study, which is intrinsic to the MRF process. Robust Equivariant Imaging (REI)~\cite{ref:chen2022} has recently been proposed which directly addresses noise within scans. We hope to extend our work to REI to address this drawback.

NLEI succeeds over EI by addressing the nonlinearity of the MRF inverse problem which exploits additional temporal-domain priors, while EI only takes advantage of the spatial-domain image priors obtained form transformation invariances. 
Supervised learning (RCA-U-Net) provides the best performance, with self-supervised learning (NLEI) closely approaching. The advantages of using only fast MRF scans for training, no ground truth requirement, and flexibility across pathologies which may not be available in the ground truth, may make NLEI a competitive alternative.

\section{Conclusion}
\label{sec:conclusion}

A proof-of-concept for a self-supervised learning approach (NLEI) for MRF multi-parametric quantitative tissue mapping was proposed. 
The method was validated on two cartesian (FFT) k-space sampling patterns on retrospectively simulated MRF data. The NLEI's performance was observed to approach the state-of-the-art supervised learning methods, despite not using ground truth for training. Future work will include extensions to address noisy, non-cartesian acquisitions from prospective in-vivo scans.

\section{Compliance with Ethical Standards}
\label{sec:ethics}

This research study was conducted retrospectively using anonymised human subject scans made available by GE Healthcare who obtained informed consent in compliance with the German Act on Medical Devices. Approval was granted by the Ethics Committee of The University of Bath (Date. Sept 2021 / No. 6568).

\section{Acknowledgments}
\label{sec:acknowledgments}

CMP and MIM are supported by the EU’s Horizon 2020 (grant No. 952172). MG is supported by the EPSRC grant EP/X001091/1.

\bibliographystyle{IEEEbib}
\bibliography{refs}

\end{document}


\ninept
\maketitle

\section{Data Acquired}
\label{sec:sup_dataset}
Data was obtained from a 3T GE scanner (MR750w system - GE Healthcare, Waukesha, WI) with 8-channel receive-only head RF coil, $230\times230$ mm$^2$ FOV, $5$ mm slice thickness, and used a FISP-MRF acquisition protocol with $T=1000$ repetitions, same flip angles as~\cite{jiang2015mr}, inversion time 18 ms, repetition time 10 ms and echo time 1.8 ms. QMaps were reconstructed using the LRTV algorithm~\cite{ref:golbabaee2021-lrtv_mrfresnet} with regularisation parameter $\lambda=0$. The ground truth (reference) and simulated MRF datasets were then generated as described in the main paper.

\section{Evaluation Metrics - Extended}
\label{sec:sup_eval_metrics}
For QMaps, the Mean Absolute Error (MAE), Mean Absolute Percentage Error (MAPE), Peak Signal-to-Noise-Ratio (PSNR) and Structural Similarity Index Measure (SSIM) were used. The magnitude of $\text{PD}_{\text{complex}}$ normalised between 0 and 1 was used for evaluation. For TSMIs, the PSNR and SSIM were used. The data ranges used for PSNR and SSIM were 6s for T1, 4s for T2, 1 for PD, and the arithmetic range of each TSMI. Head-masks were applied on all reconstructions and the metrics were then calculated and averaged across 15 test slices.

\section{Searching for Optimal Values of $\alpha$}
\label{sec:sup_search_alpha}
We performed a number of experiments to determine the optimal values of the regularisation constant for the Equivariance  Imaging (EI) loss, $\alpha$, for Non-Linear Equivariant Imaging (NLEI) and linear Equivariant Imaging (EI). The same methodology was used for both NLEI and EI, as described in the main paper. The performance of NLEI and EI was tested with Spiral and EPI subsampling patterns using $\alpha$ values of $0$, Measurement Consistency (MC) only, and $10^{\{-12, -11, -10, -9, -8, -7, -6, -5, -4, -3 ,-2, -1, 0 ,1, 2, 3, 4, 5, 6\}}$. The results presented in Section.~\ref{ssec:sup_grouped_metrics} are averaged over 15 test slices. The results presented in Section.~\ref{ssec:sup_alpha_selection} are the results presented in Section.~\ref{ssec:sup_grouped_metrics} averaged using all QMaps, T1, T2 and PD. All results were masked using a head mask as mentioned in the main paper.

\subsection{Results Grouped by Metric}
\label{ssec:sup_grouped_metrics}
\begin{figure*}[h]

	\begin{minipage}[b]{17.5cm}
  		\centering
  		\centerline{\includegraphics[scale=0.4]{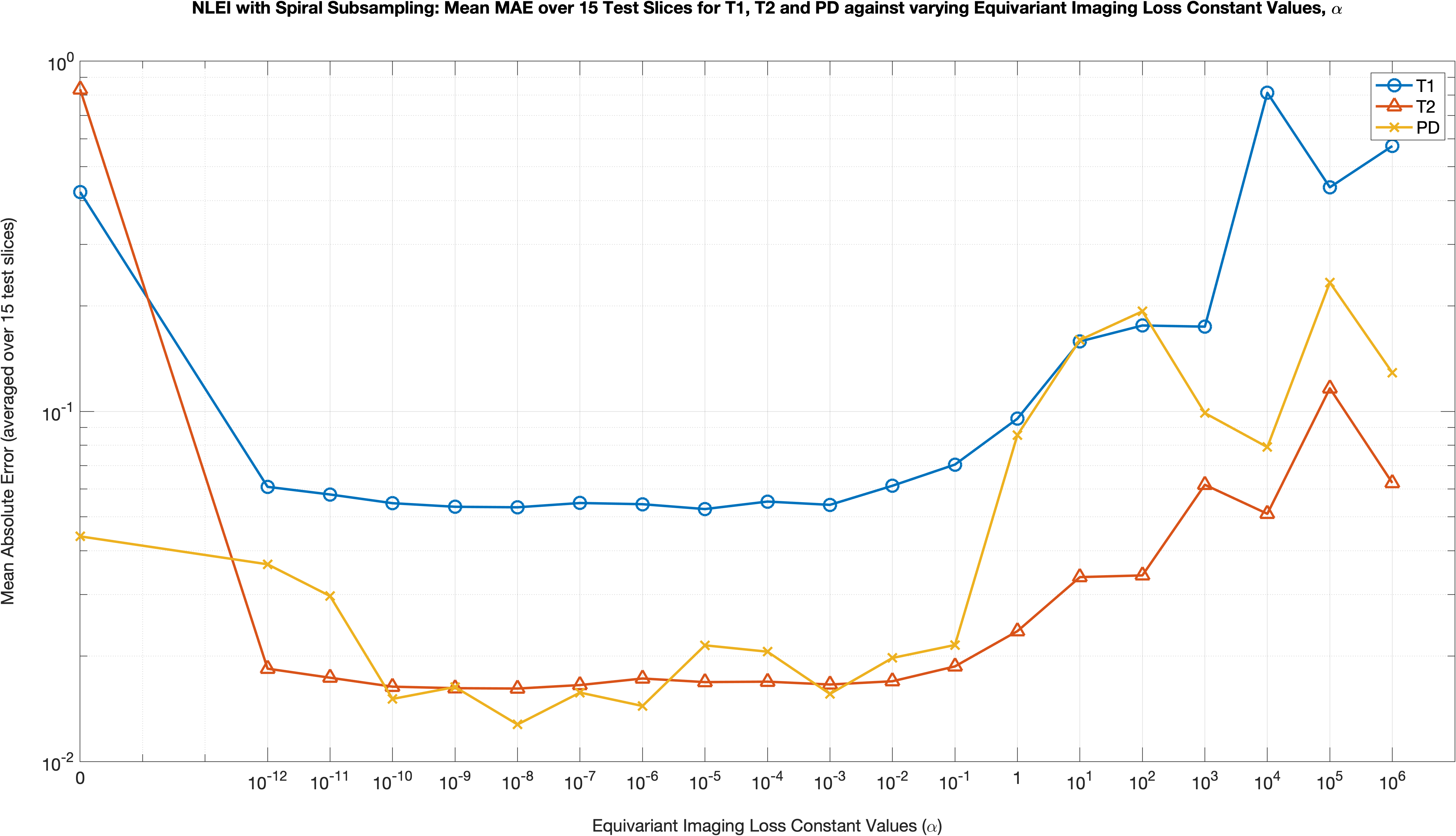}}
	\end{minipage}
	
	\caption{The graph for the average MAE across 15 test slices for each QMap, T1, T2 and PD, against the Equivariant loss hyperparameter, $\alpha$, for NonLinear Equivariant Imaging (NLEI) with Spiral Subsampling.}
	\label{fig:sup_nl_ei_spiral_fig_mae}

\end{figure*}

\begin{figure*}[h]

	\begin{minipage}[b]{17.5cm}
  		\centering
  		\centerline{\includegraphics[scale=0.4]{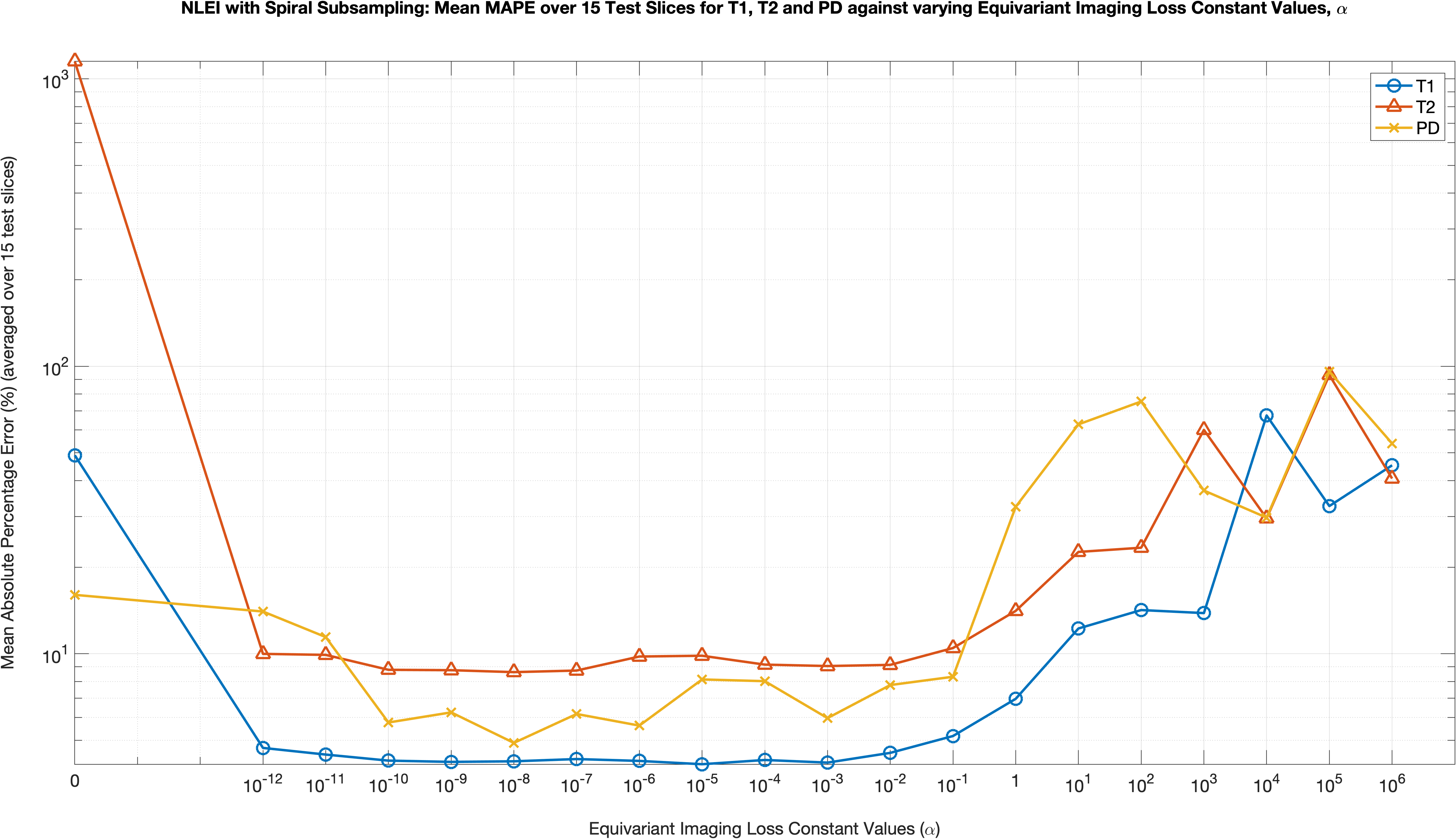}}
	\end{minipage}

	\caption{The graph for the average MAPE across 15 test slices for each QMap, T1, T2 and PD, against the Equivariant loss hyperparameter, $\alpha$, for NonLinear Equivariant Imaging (NLEI) with Spiral Subsampling.}
	\label{fig:sup_nl_ei_spiral_fig_mape}

\end{figure*}

\begin{figure*}[h]

	\begin{minipage}[b]{17.5cm}
  		\centering
  		\centerline{\includegraphics[scale=0.4]{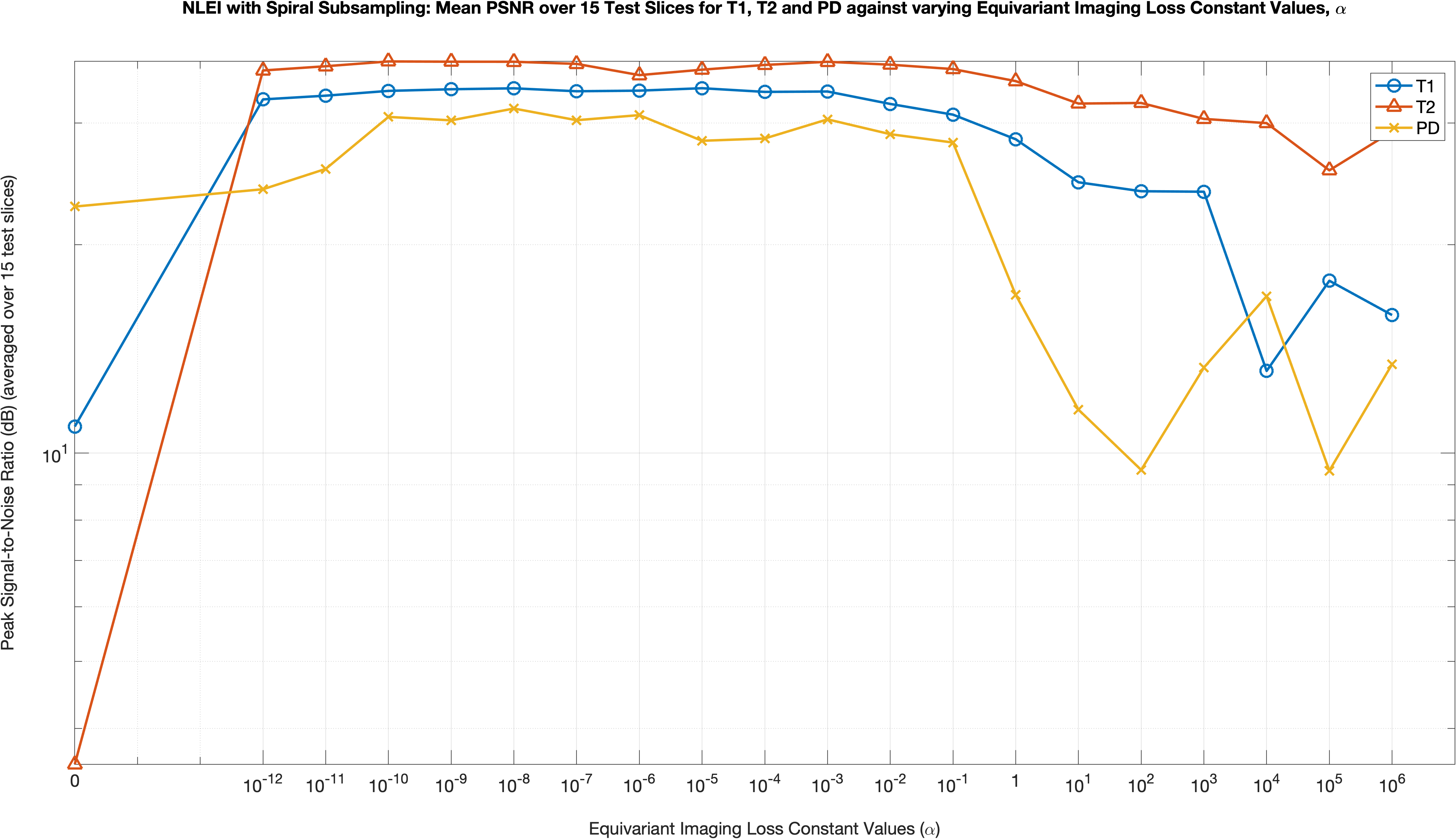}}
	\end{minipage}

	\caption{The graph for the average PSNR across 15 test slices for each QMap, T1, T2 and PD, against the Equivariant loss hyperparameter, $\alpha$, for NonLinear Equivariant Imaging (NLEI) with Spiral Subsampling.}
	\label{fig:sup_nl_ei_spiral_fig_psnr}

\end{figure*}

\begin{figure*}[h]

	\begin{minipage}[b]{17.5cm}
  		\centering
  		\centerline{\includegraphics[scale=0.4]{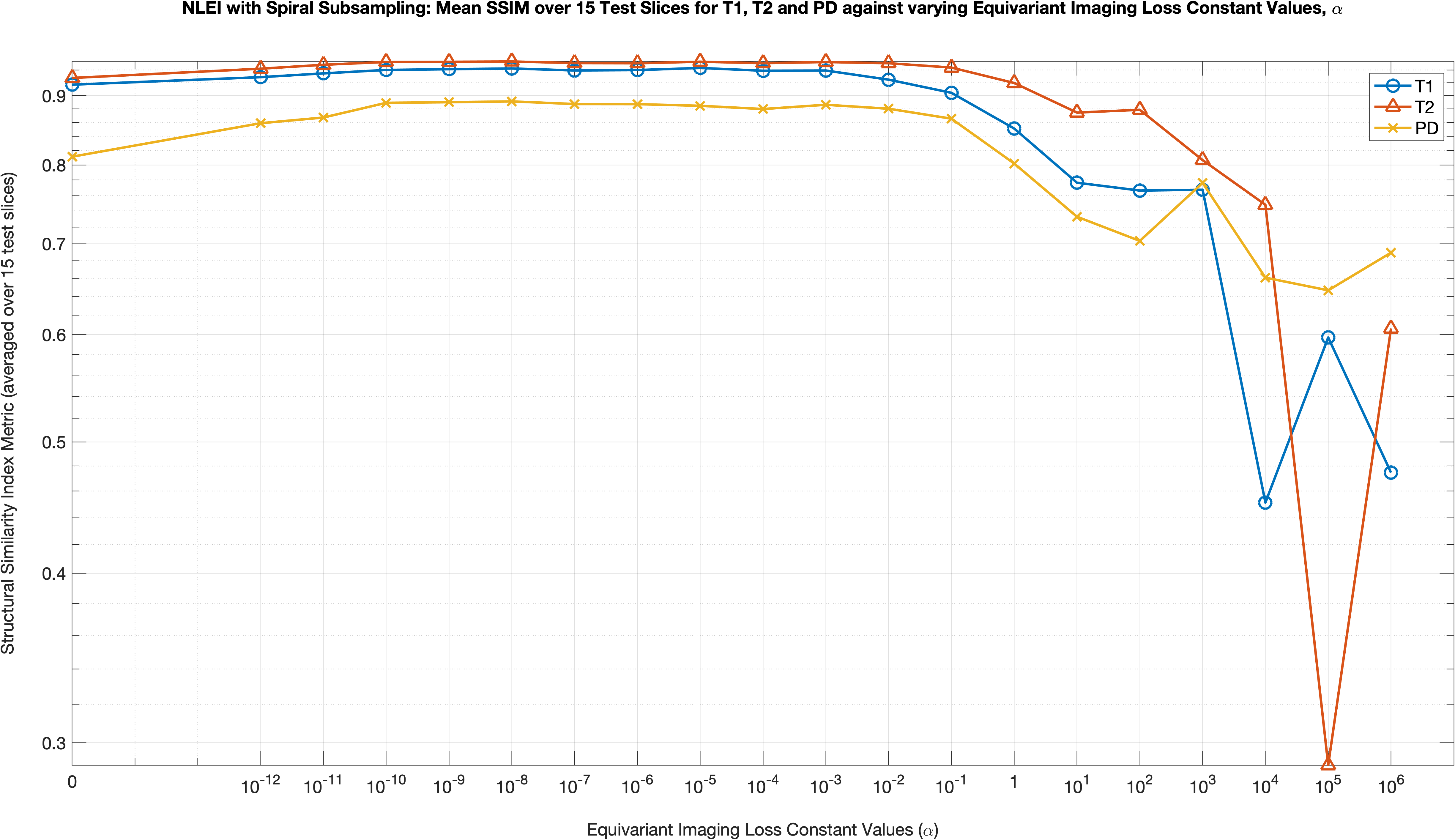}}
	\end{minipage}

	\caption{The graph for the average SSIM across 15 test slices for each QMap, T1, T2 and PD, against the Equivariant loss hyperparameter, $\alpha$, for NonLinear Equivariant Imaging (NLEI) with Spiral Subsampling.}
	\label{fig:sup_nl_ei_spiral_fig_ssim}

\end{figure*}

\begin{figure*}[h]

	\begin{minipage}[b]{17.5cm}
  		\centering
  		\centerline{\includegraphics[scale=0.4]{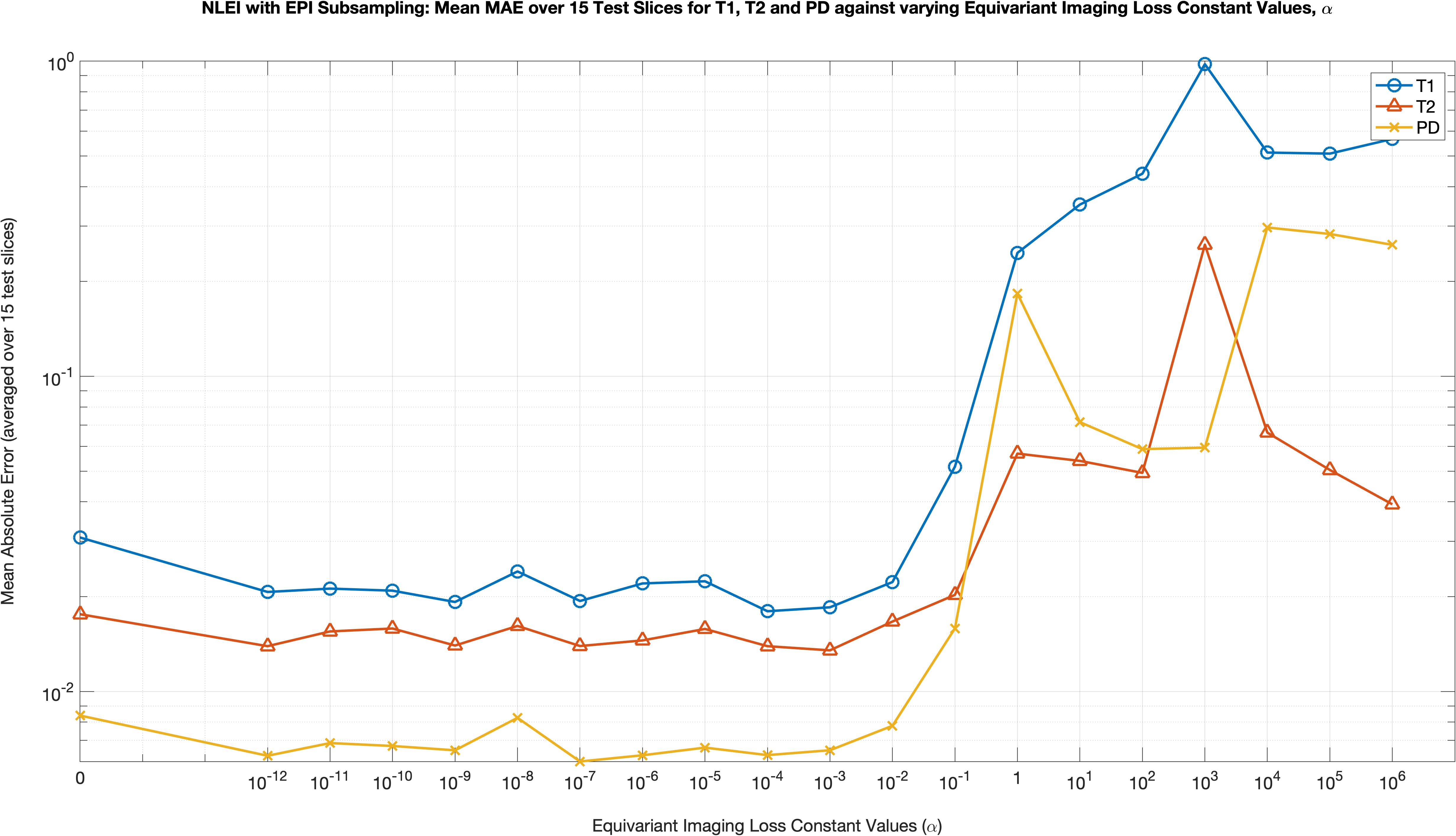}}
	\end{minipage}
	
	\caption{The graph for the average MAE across 15 test slices for each QMap, T1, T2 and PD, against the Equivariant loss hyperparameter, $\alpha$, for NonLinear Equivariant Imaging (NLEI) with EPI Subsampling.}
	\label{fig:sup_nl_ei_epi_fig_mae}

\end{figure*}

\begin{figure*}[h]

	\begin{minipage}[b]{17.5cm}
  		\centering
  		\centerline{\includegraphics[scale=0.4]{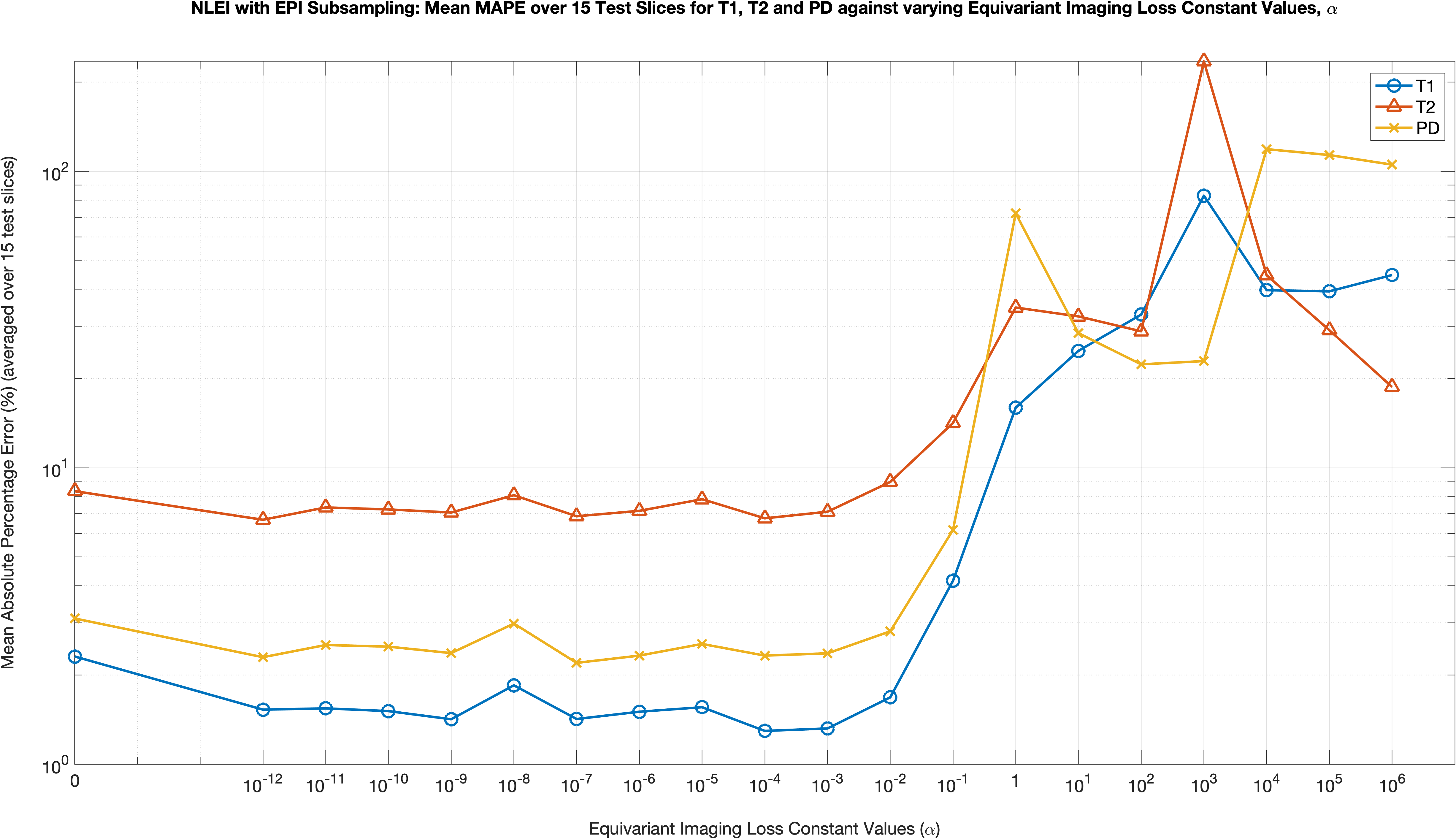}}
	\end{minipage}
	
	\caption{The graph for the average MAPE across 15 test slices for each QMap, T1, T2 and PD, against the Equivariant loss hyperparameter, $\alpha$, for NonLinear Equivariant Imaging (NLEI) with EPI Subsampling.}
	\label{fig:sup_nl_ei_epi_fig_mape}

\end{figure*}

\begin{figure*}[h]

	\begin{minipage}[b]{17.5cm}
  		\centering
  		\centerline{\includegraphics[scale=0.4]{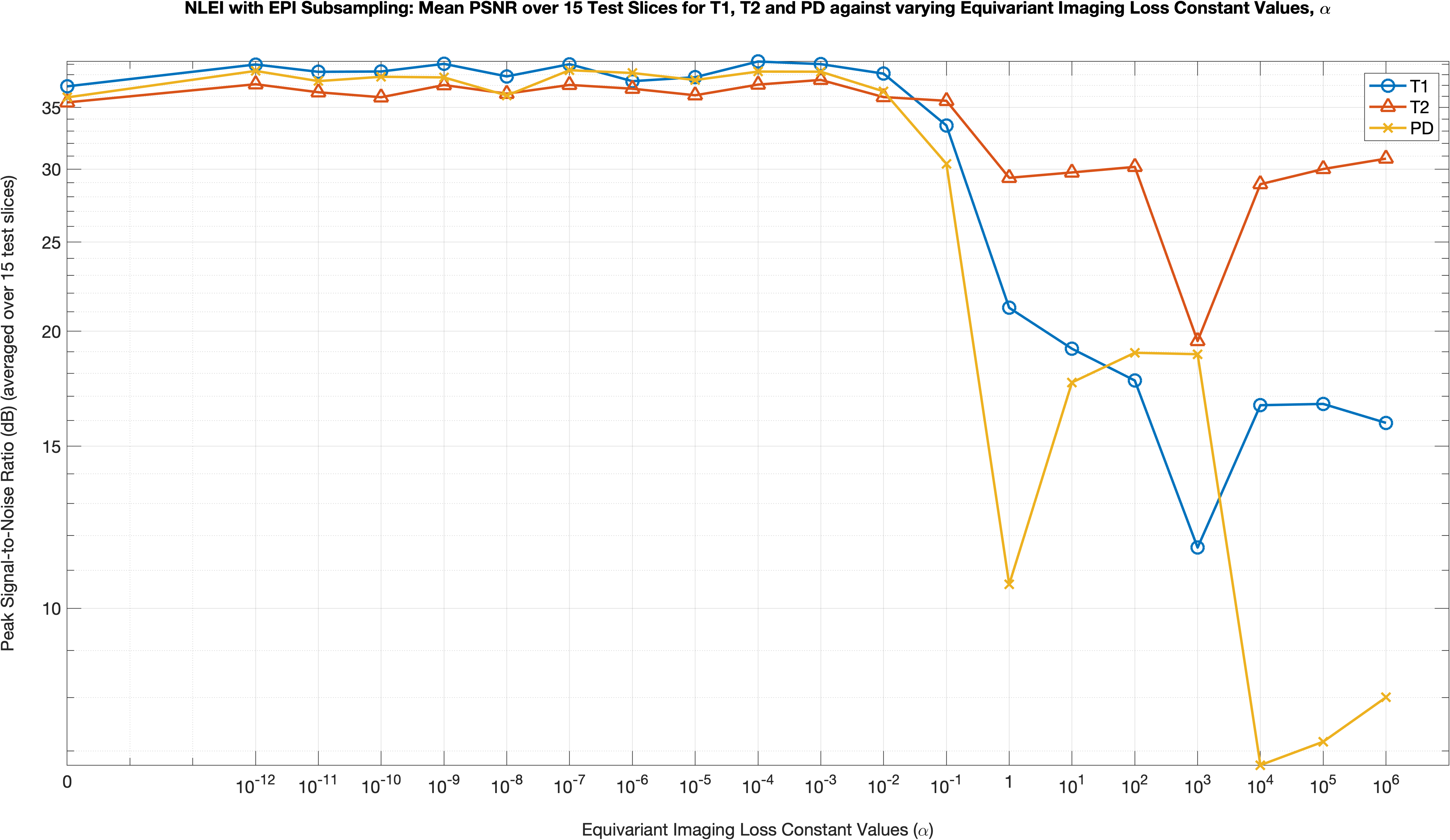}}
	\end{minipage}

	\caption{The graph for the average PSNR across 15 test slices for each QMap, T1, T2 and PD, against the Equivariant loss hyperparameter, $\alpha$, for NonLinear Equivariant Imaging (NLEI) with EPI Subsampling.}
	\label{fig:sup_nl_ei_epi_fig_psnr}

\end{figure*}

\begin{figure*}[h]

	\begin{minipage}[b]{17.5cm}
  		\centering
  		\centerline{\includegraphics[scale=0.4]{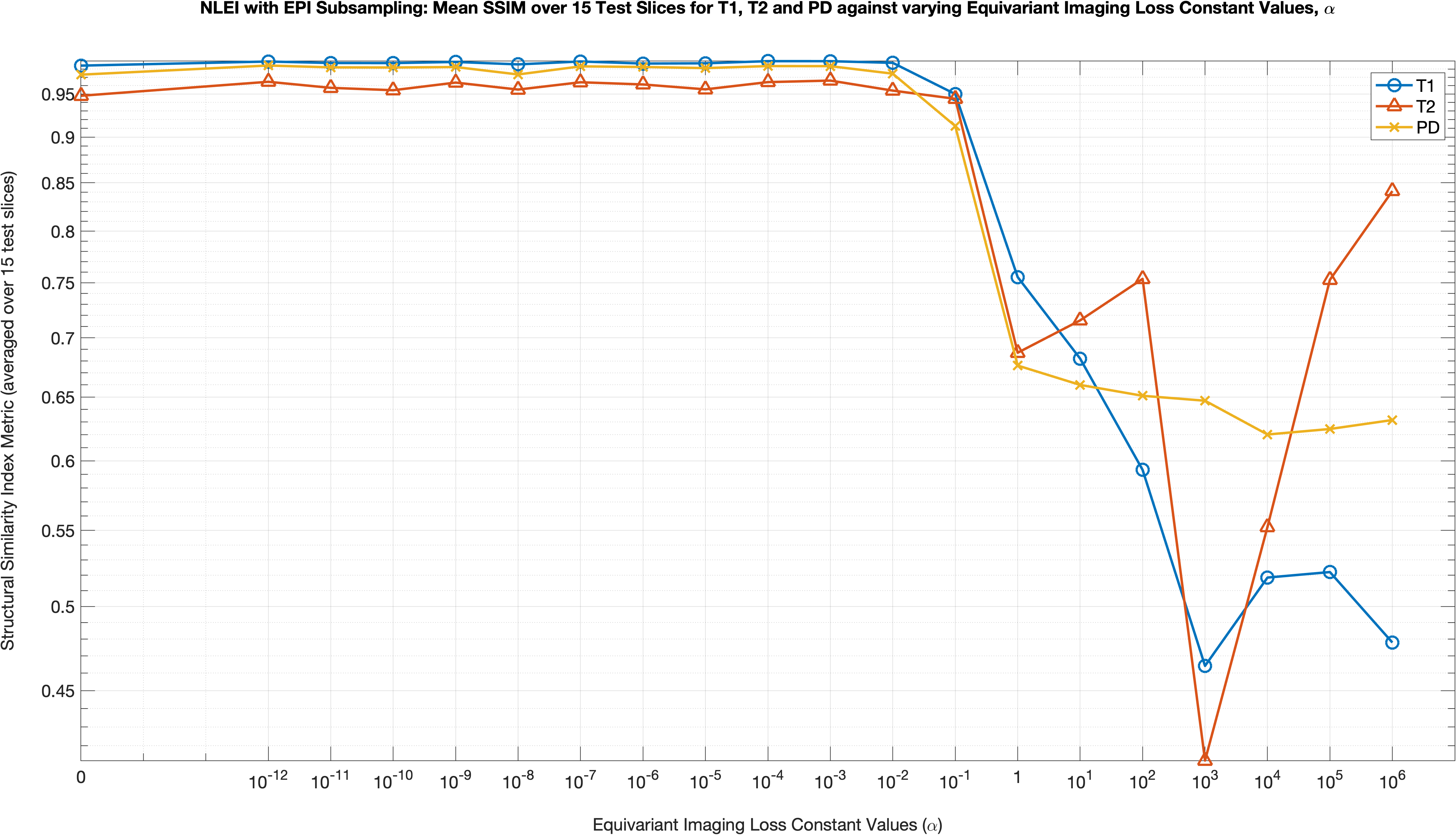}}
	\end{minipage}

	\caption{The graph for the average SSIM across 15 test slices for each QMap, T1, T2 and PD, against the Equivariant loss hyperparameter, $\alpha$, for NonLinear Equivariant Imaging (NLEI) with EPI Subsampling.}
	\label{fig:sup_nl_ei_epi_fig_ssim}

\end{figure*}

\begin{figure*}[h]

	\begin{minipage}[b]{17.5cm}
  		\centering
  		\centerline{\includegraphics[scale=0.4]{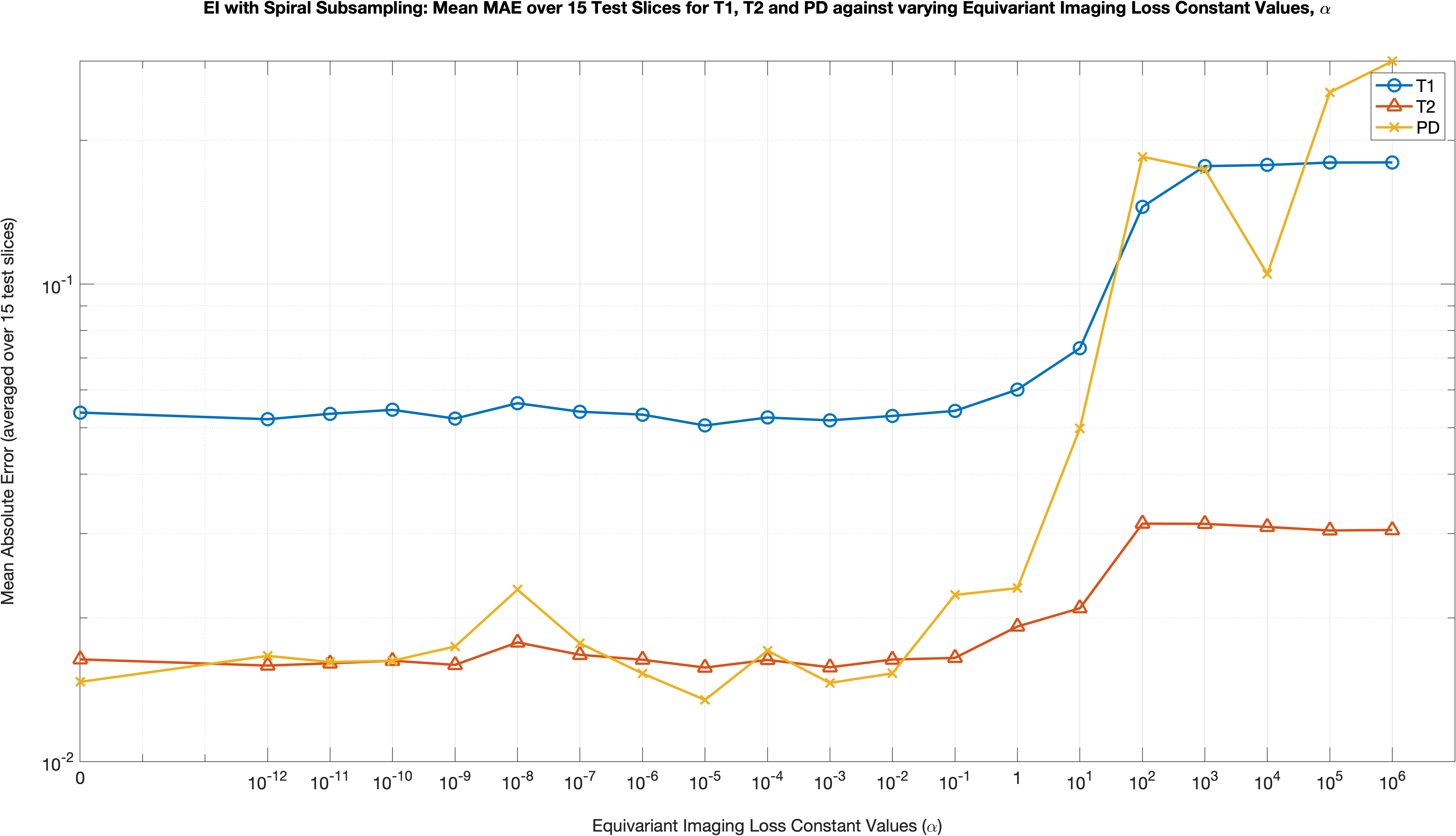}}
	\end{minipage}

	\caption{The graph for the average MAE across 15 test slices for each QMap, T1, T2 and PD, against the Equivariant loss hyperparameter, $\alpha$, for linear Equivariant Imaging (EI) with Spiral Subsampling.}
	\label{fig:sup_ei_spiral_fig_mae}

\end{figure*}

\begin{figure*}[h]

	\begin{minipage}[b]{17.5cm}
  		\centering
  		\centerline{\includegraphics[scale=0.4]{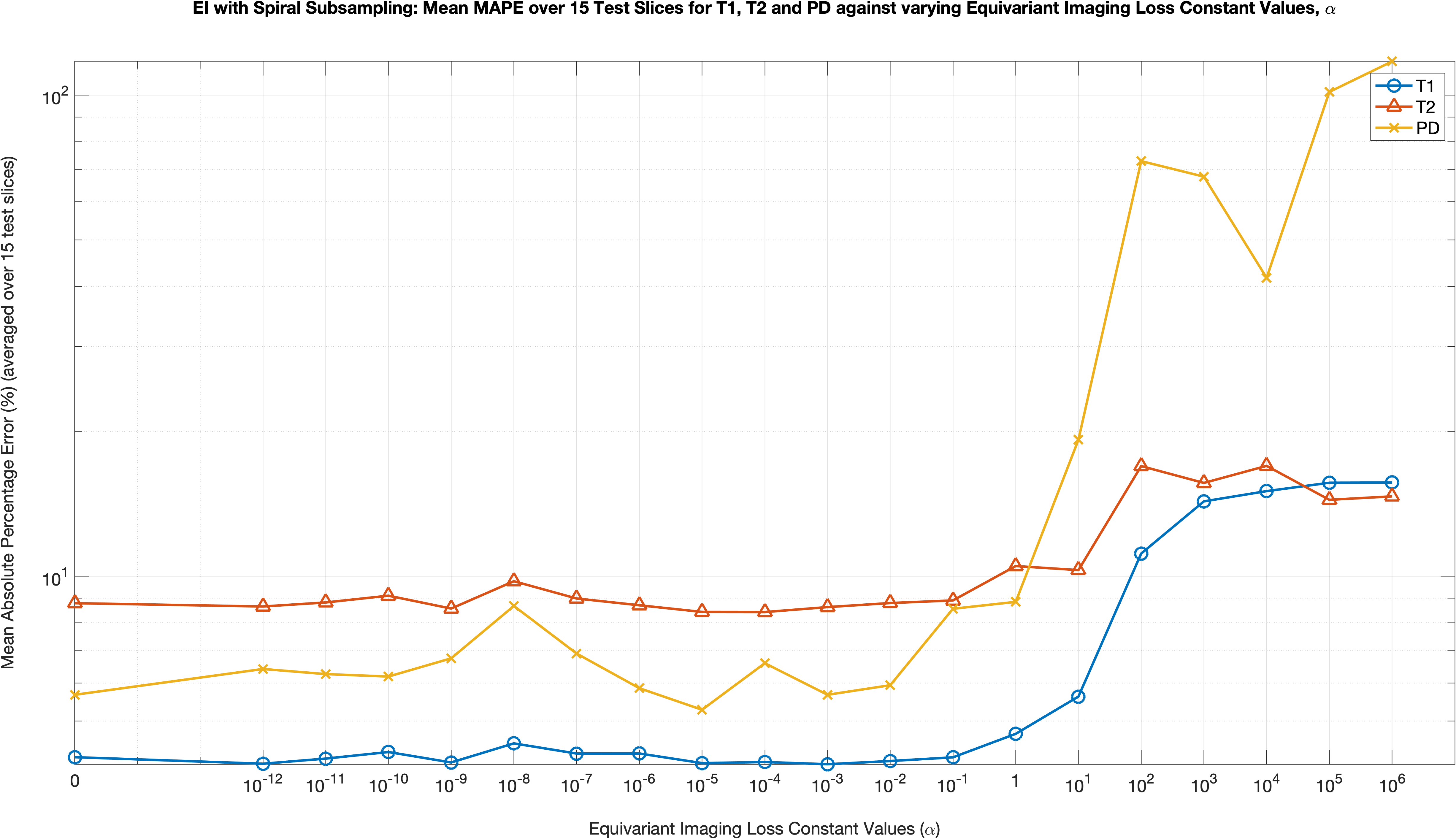}} 
	\end{minipage}

	\caption{The graph for the average MAPE across 15 test slices for each QMap, T1, T2 and PD, against the Equivariant loss hyperparameter, $\alpha$, for linear Equivariant Imaging (EI) with Spiral Subsampling.}
	\label{fig:sup_ei_spiral_fig_mape}

\end{figure*}

\begin{figure*}[h]

	\begin{minipage}[b]{17.5cm}
  		\centering
  		\centerline{\includegraphics[scale=0.4]{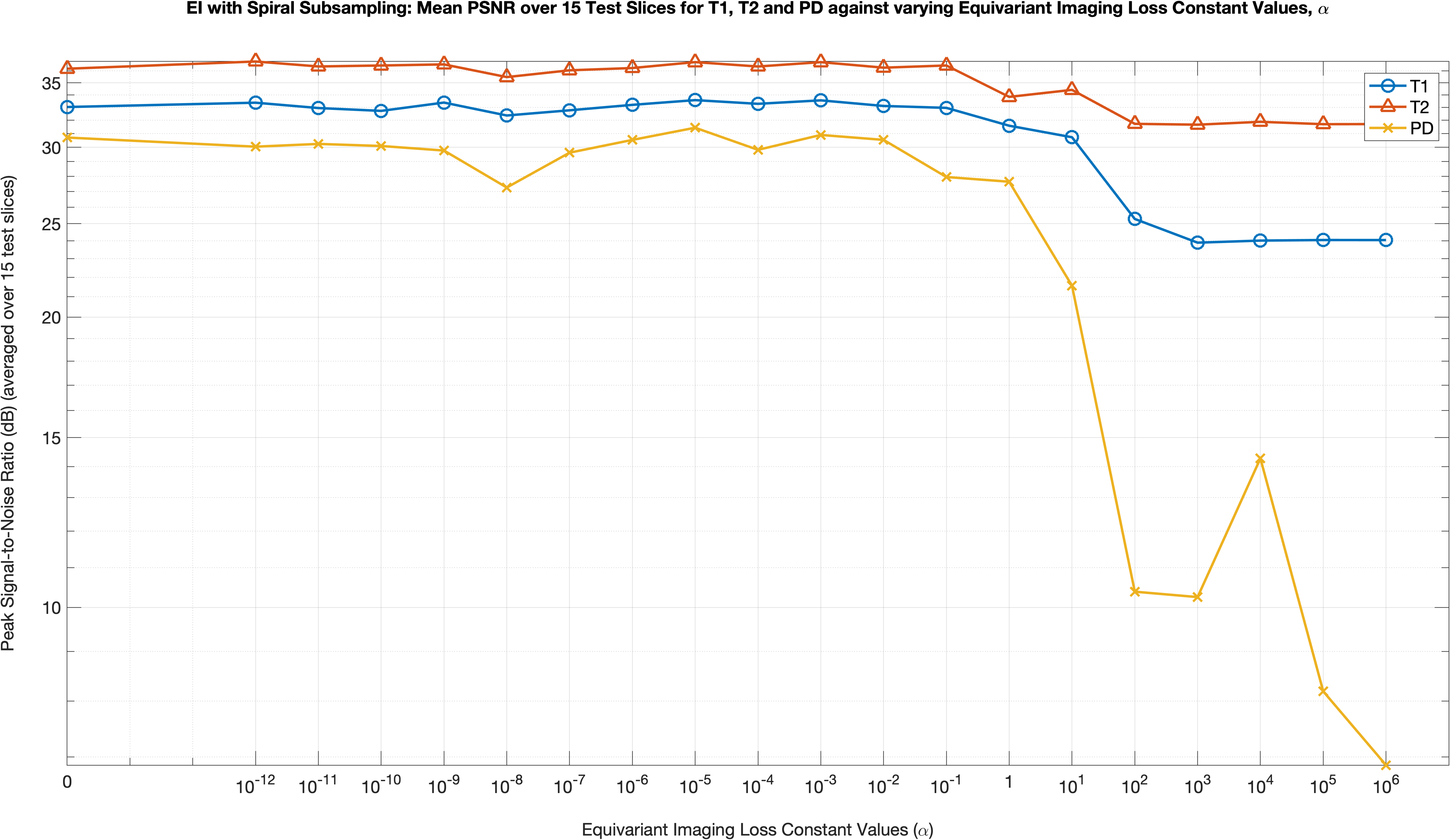}}
	\end{minipage}

	\caption{The graph for the average PSNR across 15 test slices for each QMap, T1, T2 and PD, against the Equivariant loss hyperparameter, $\alpha$, for linear Equivariant Imaging (EI) with Spiral Subsampling.}
	\label{fig:sup_ei_spiral_fig_psnr}

\end{figure*}

\begin{figure*}[h]

	\begin{minipage}[b]{17.5cm}
 		\centering
  		\centerline{\includegraphics[scale=0.4]{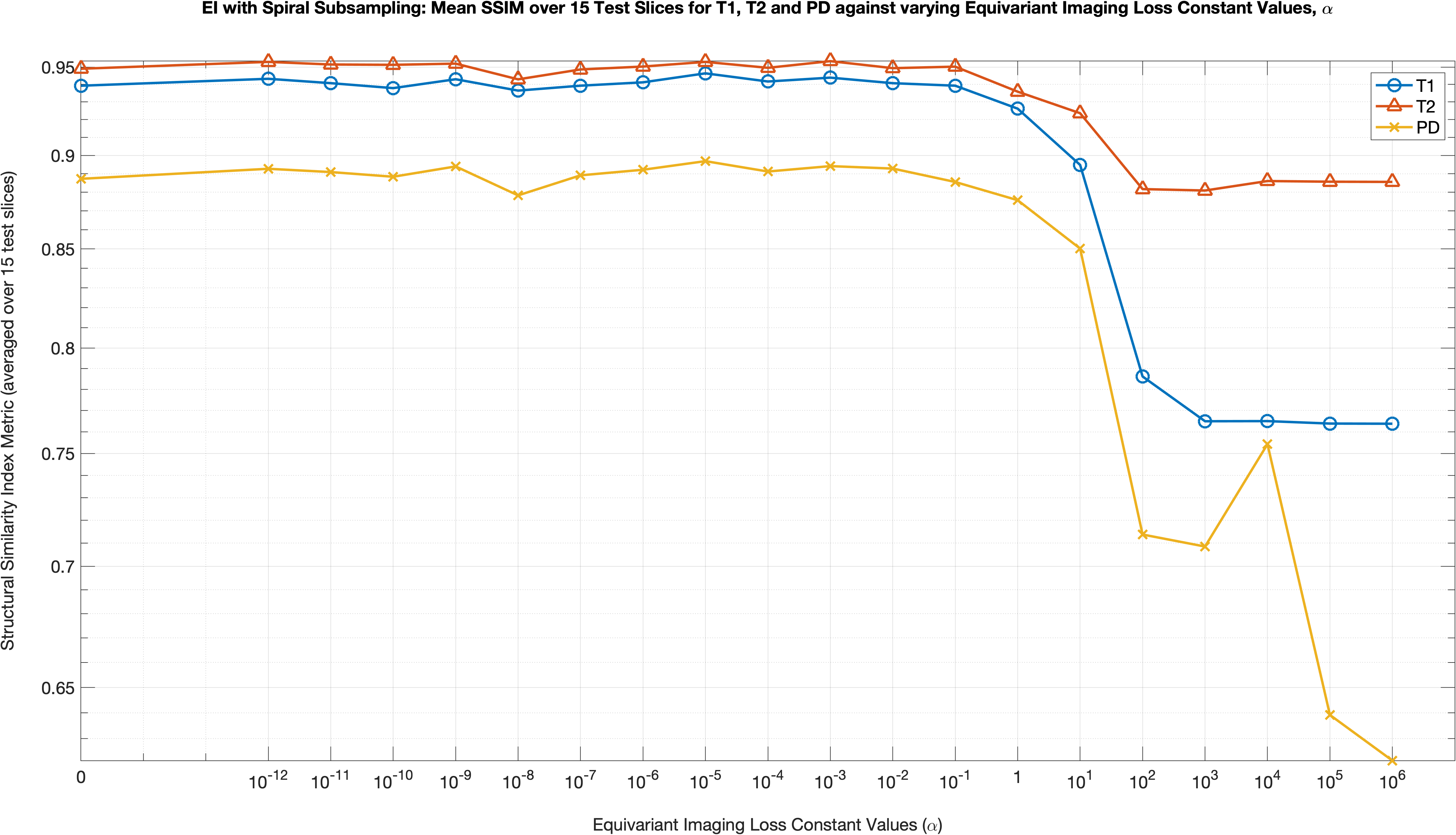}}
	\end{minipage}

	\caption{The graph for the average SSIM across 15 test slices for each QMap, T1, T2 and PD, against the Equivariant loss hyperparameter, $\alpha$, for linear Equivariant Imaging (EI) with Spiral Subsampling.}
	\label{fig:sup_ei_spiral_fig_ssim}

\end{figure*}

\begin{figure*}[h]

	\begin{minipage}[b]{17.5cm}
		\centering
  		\centerline{\includegraphics[scale=0.4]{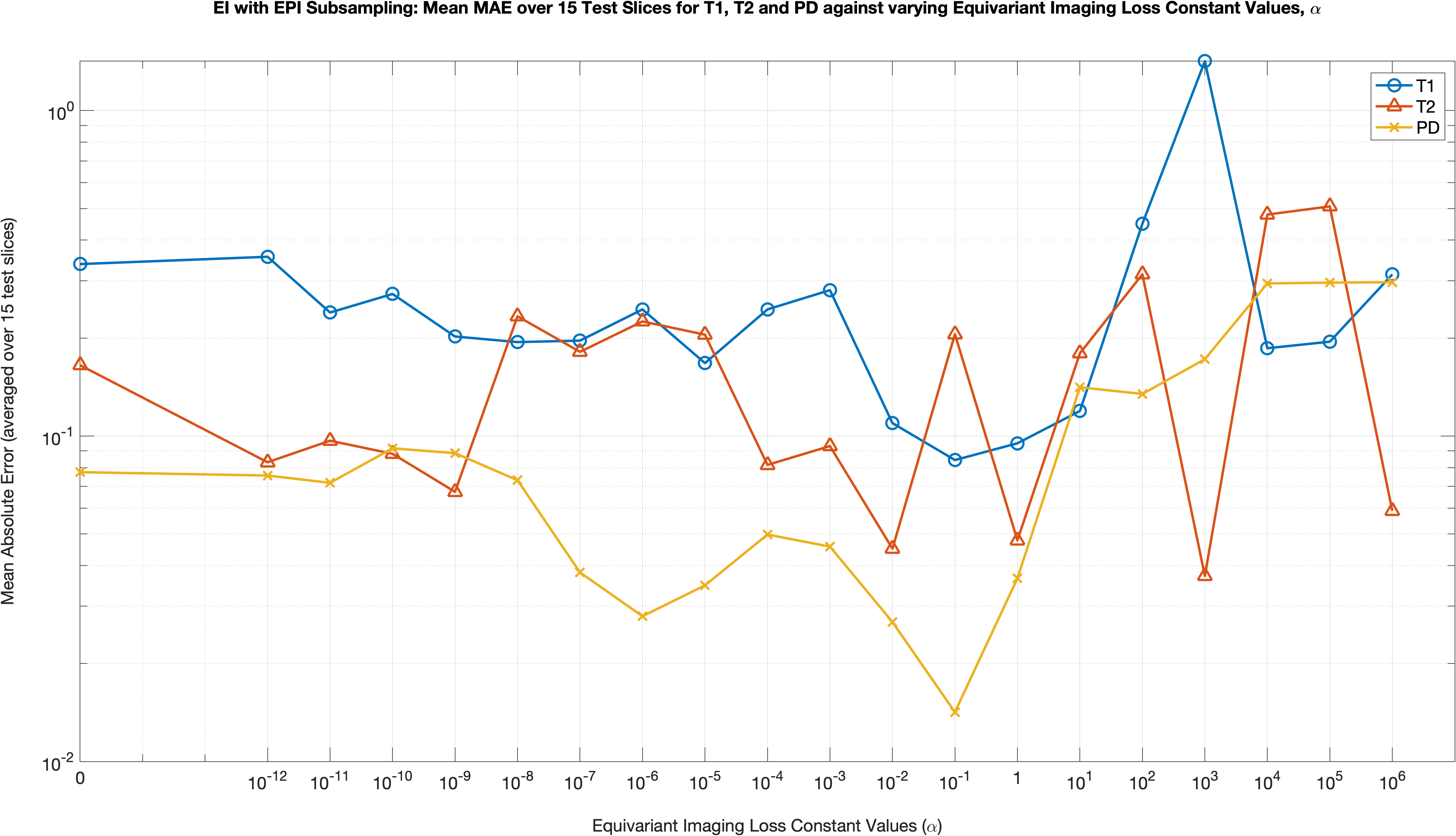}}
	\end{minipage}

	\caption{The graph for the average MAE across 15 test slices for each QMap, T1, T2 and PD, against the Equivariant loss hyperparameter, $\alpha$, for linear Equivariant Imaging (EI) with EPI Subsampling.}
	\label{fig:sup_ei_epi_fig_mae}

\end{figure*}

\begin{figure*}[h]

	\begin{minipage}[b]{17.5cm}
  		\centering
  		\centerline{\includegraphics[scale=0.4]{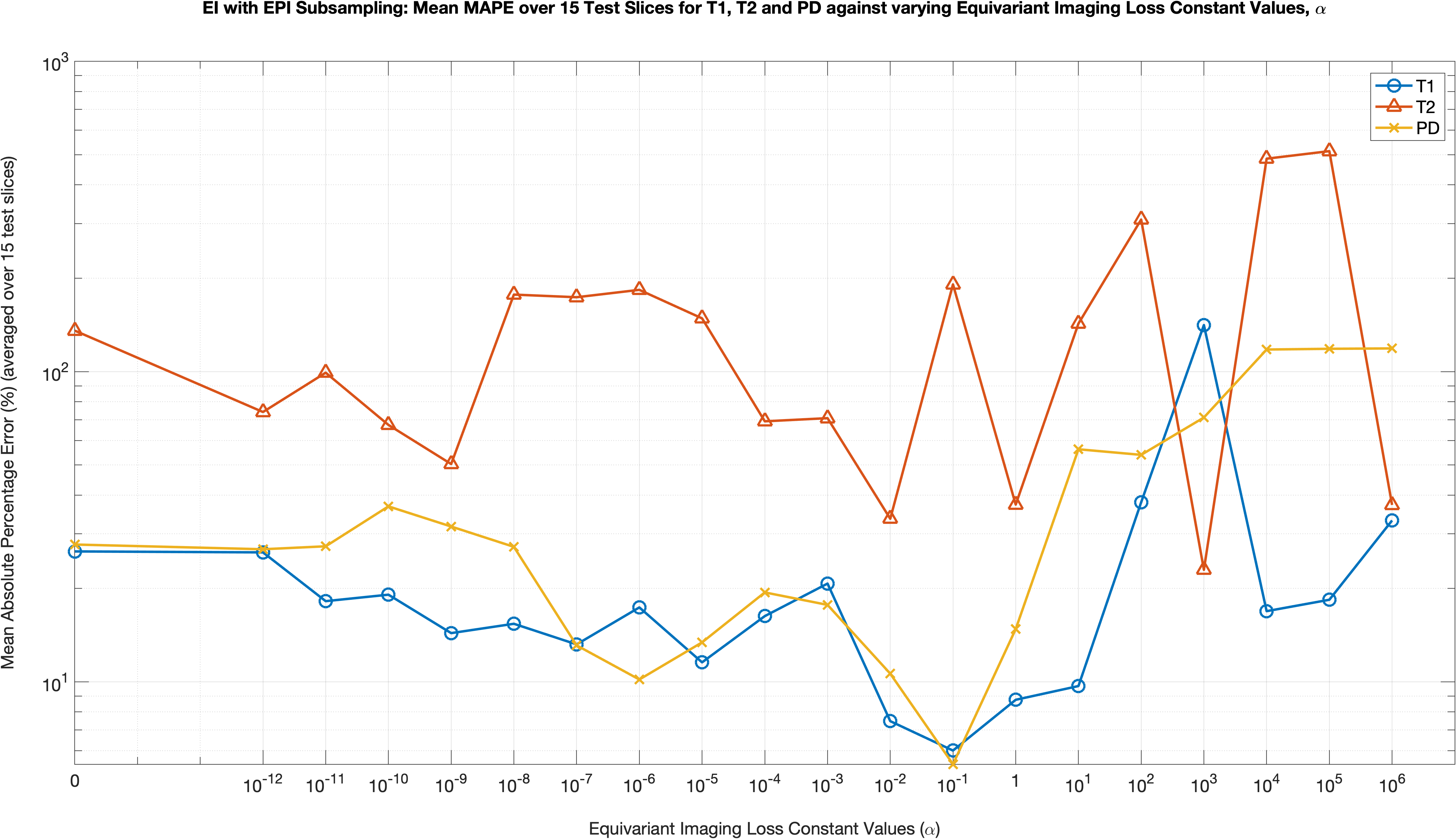}}
	\end{minipage}

	\caption{The graph for the average MAPE across 15 test slices for each QMap, T1, T2 and PD, against the Equivariant loss hyperparameter, $\alpha$, for linear Equivariant Imaging (EI) with EPI Subsampling.}
	\label{fig:sup_ei_epi_fig_mape}

\end{figure*}

\begin{figure*}[h]

	\begin{minipage}[b]{17.5cm}
  		\centering
  		\centerline{\includegraphics[scale=0.4]{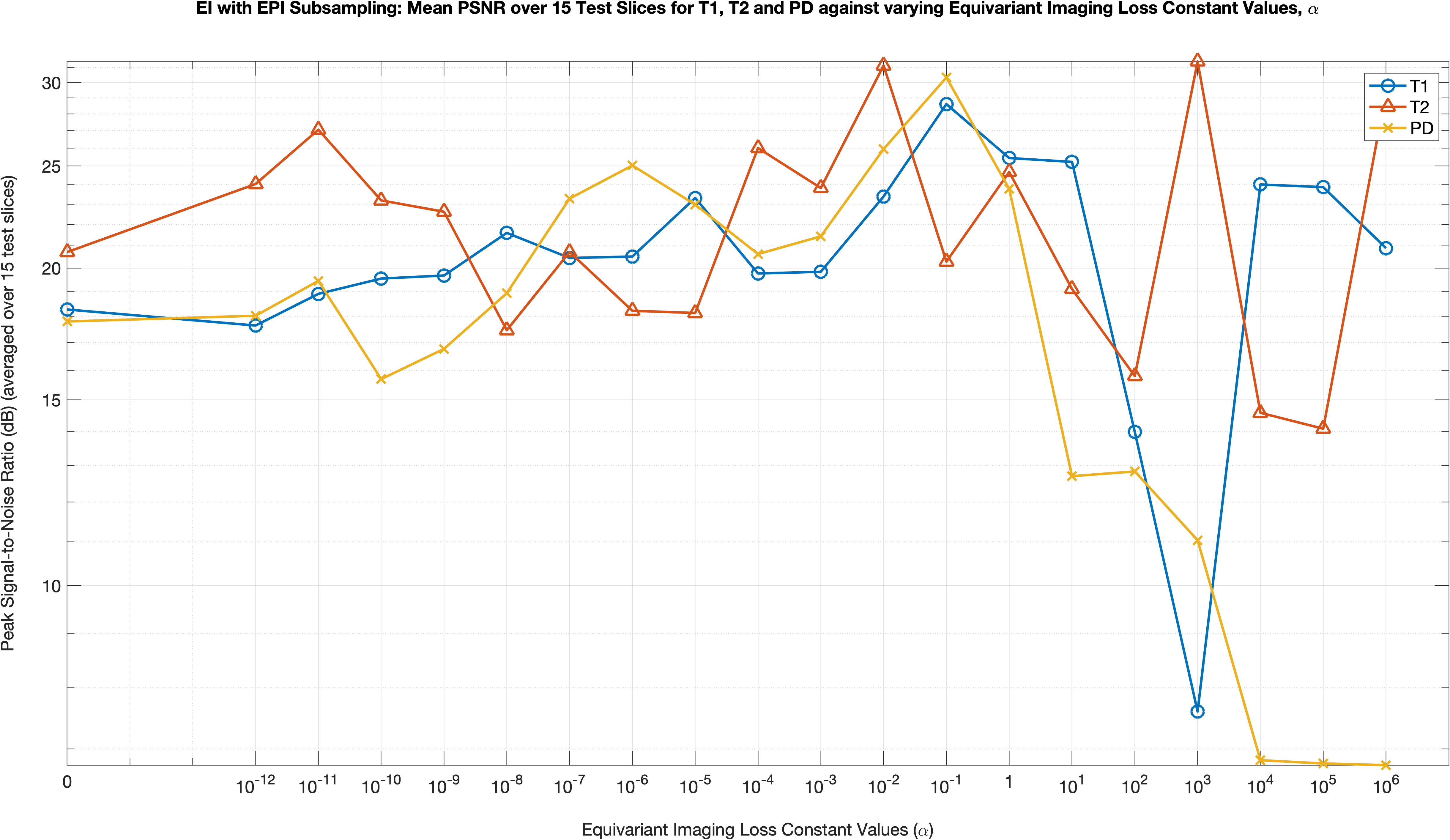}}
	\end{minipage}

	\caption{The graph for the average PSNR across 15 test slices for each QMap, T1, T2 and PD, against the Equivariant loss hyperparameter, $\alpha$, for linear Equivariant Imaging (EI) with EPI Subsampling.}
	\label{fig:sup_ei_epi_fig_psnr}

\end{figure*}

\begin{figure*}[h]

	\begin{minipage}[b]{17.5cm}
  		\centering
  		\centerline{\includegraphics[scale=0.4]{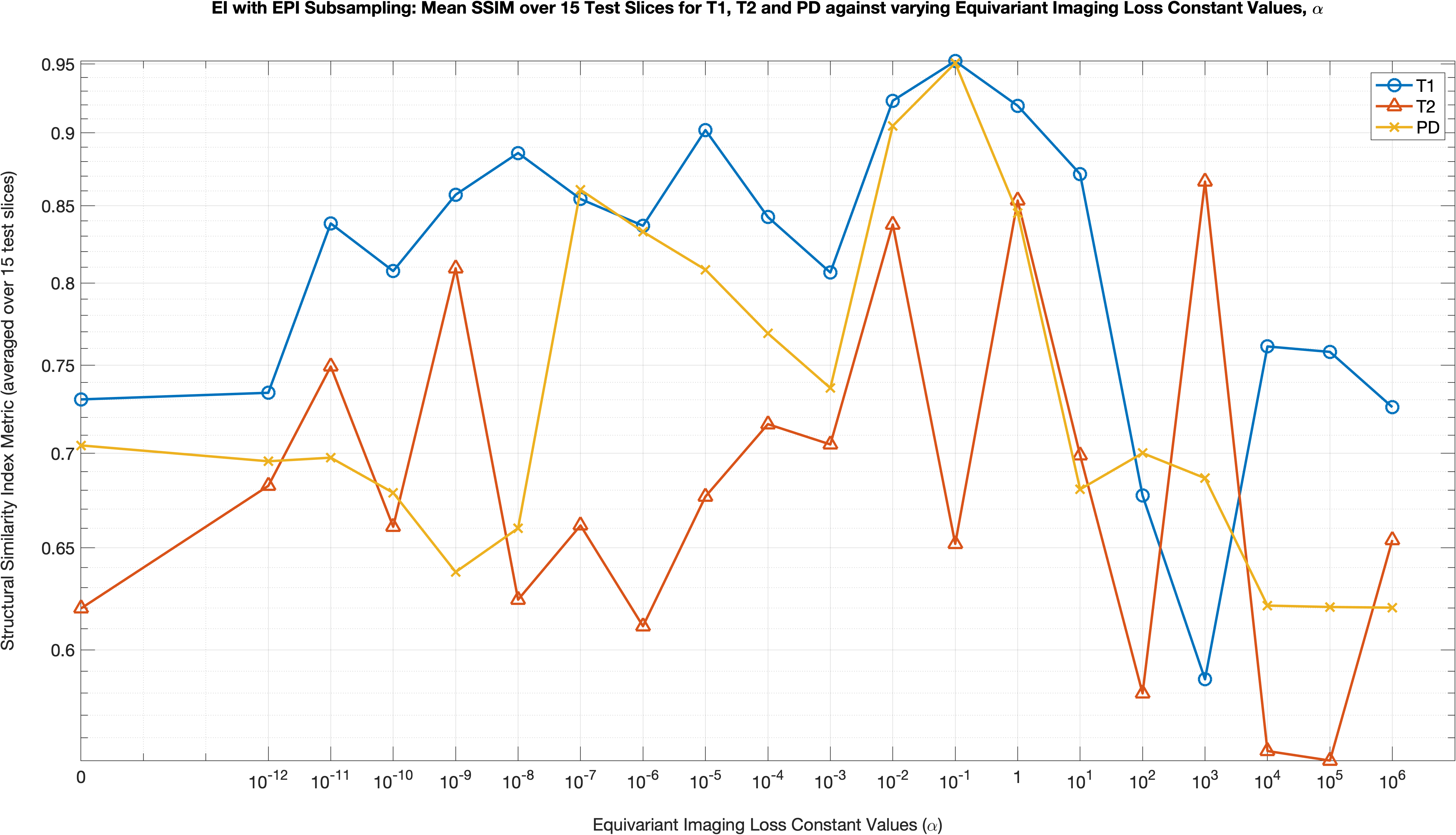}}
	\end{minipage}

	\caption{The graph for the average SSIM across 15 test slices for each QMap, T1, T2 and PD, against the Equivariant loss hyperparameter, $\alpha$, for linear Equivariant Imaging (EI) with EPI Subsampling.}
	\label{fig:sup_ei_epi_fig_ssim}

\end{figure*}

We initially group the Time-Series Magnetisation Images (TSMIs) and Quantitative Maps (QMaps) (T1, T2 and Proton Density, PD) results by metric. We find a range of values for $\alpha$ which give the best performance with respect to particular metrics and QMaps, see \Cref{tab:sup_nl_ei_spiral_table_mae,tab:sup_nl_ei_spiral_table_mape,tab:sup_nl_ei_spiral_table_psnr,tab:sup_nl_ei_spiral_table_ssim,tab:sup_nl_ei_epi_table_mae,tab:sup_nl_ei_epi_table_mape,tab:sup_nl_ei_epi_table_psnr,tab:sup_nl_ei_epi_table_ssim} and \Cref{fig:sup_nl_ei_spiral_fig_mae,fig:sup_nl_ei_spiral_fig_mape,fig:sup_nl_ei_spiral_fig_psnr,fig:sup_nl_ei_spiral_fig_ssim,fig:sup_nl_ei_epi_fig_mae,fig:sup_nl_ei_epi_fig_mape,fig:sup_nl_ei_epi_fig_psnr,fig:sup_nl_ei_epi_fig_ssim} for NLEI, and \Cref{tab:sup_ei_spiral_table_mae,tab:sup_ei_spiral_table_mape,tab:sup_ei_spiral_table_psnr,tab:sup_ei_spiral_table_ssim,tab:sup_ei_epi_table_mae,tab:sup_ei_epi_table_mape,tab:sup_ei_epi_table_psnr,tab:sup_ei_epi_table_ssim} and \Cref{fig:sup_ei_spiral_fig_mae,fig:sup_ei_spiral_fig_mape,fig:sup_ei_spiral_fig_psnr,fig:sup_ei_spiral_fig_ssim,fig:sup_ei_epi_fig_mae,fig:sup_ei_epi_fig_mape,fig:sup_ei_epi_fig_psnr,fig:sup_ei_epi_fig_ssim} for EI. 

\subsection{Optimal $\alpha$ Selection}
\label{ssec:sup_alpha_selection}
To determine the optimal value of $\alpha$, the mean of all QMap results for each metric, for each value of $\alpha$ is taken, e.g for $\alpha=0$ and NLEI with Spiral subsampling, the mean is taken using the MAE results for T1, T2 and PD, then the mean is taken using the MAPE results for T1, T2 and PD, and similarly for PSNR and SSIM. We then determined the best averaged QMaps value of each metric against $\alpha$, see \Cref{tab:sup_nl_ei_spiral_table_avg_qmaps,tab:sup_nl_ei_epi_table_avg_qmaps,tab:sup_ei_spiral_table_avg_qmaps,tab:sup_ei_epi_table_avg_qmaps}. The optimal value of $\alpha$ was then chosen according to two criteria: (i) the number of metrics which agree to a particular value of $\alpha$ and (ii) according to metric importance given by the order 1) MAPE, 2) MAE, 3) PSNR and 4) SSIM. We prioritise data quality metrics, MAPE and MAE, over image quality metrics, PSNR and SSIM.

We find the optimal value of $\alpha$ is dependent on the algorithm used, NLEI or EI, and the subsampling pattern used, Spiral or EPI, see \Cref{tab:sup_nl_ei_spiral_table_avg_qmaps,tab:sup_nl_ei_epi_table_avg_qmaps,tab:sup_ei_spiral_table_avg_qmaps,tab:sup_ei_epi_table_avg_qmaps}.

\bibliographystyle{IEEEbib}
\bibliography{refs}

%
\clearpage
\begin{table*}[t] 
	
	\centering
	
	\begin{tabular}{ | c | c | c | c | } 
		
		\hline \multicolumn{4}{| c |}{NLEI with Spiral Subsampling - Mean MAE} \\ \hline
		
		$\alpha$ 				& T1 MAE (s) & T2 MAE (s) & PD (a.u) \\ \hline
		 
		$\text{0}$				&	0.0538	&	0.0164	&	0.0147	\\
		$\text{10}^{\text{-12}}$	&	0.0521	&	0.0159	&	0.0166	\\
		$\text{10}^{\text{-11}}$	&	0.0535	&	0.0161	&	0.0162	\\
		$\text{10}^{\text{-10}}$	&	0.0545	&	0.0163	&	0.0163	\\
		$\text{10}^{\text{-9}}$	&	0.0523	&	0.0159	&	0.0174	\\
		$\text{10}^{\text{-8}}$	&	0.0563	&	0.0178	&	0.0229	\\
		$\text{10}^{\text{-7}}$	&	0.0540	&	0.0167	&	0.0177	\\
		$\text{10}^{\text{-6}}$	&	0.0533	&	0.0163	&	0.0153	\\
		$\text{10}^{\text{-5}}$	&	0.0506	&	0.0157	&	0.0135	\\
		$\text{10}^{\text{-4}}$	&	0.0525	&	0.0163	&	0.0171	\\
		$\text{10}^{\text{-3}}$	&	0.0518	&	0.0158	&	0.0146	\\
		$\text{10}^{\text{-2}}$	&	0.0530	&	0.0164	&	0.0153	\\
		$\text{10}^{\text{-1}}$	&	0.0543	&	0.0165	&	0.0223	\\
		$\text{1}$				&	0.0601	&	0.0192	&	0.0231	\\
		$\text{10}^{\text{1}}$	&	0.0734	&	0.0209	&	0.0499	\\
		$\text{10}^{\text{2}}$	&	0.1451	&	0.0315	&	0.1849	\\
		$\text{10}^{\text{3}}$	&	0.1767	&	0.0315	&	0.1737	\\
		$\text{10}^{\text{4}}$	&	0.1777	&	0.0310	&	0.1050	\\
		$\text{10}^{\text{5}}$	&	0.1798	&	0.0305	&	0.2523	\\
		$\text{10}^{\text{6}}$	&	0.1799	&	0.0305	&	0.2931	\\
		\hline
	
	\end{tabular}
	
	\caption[...]{The average MAE across 15 test slices for each QMap, T1, T2 and PD, against the Equivariant loss hyperparameter, $\alpha$, for NonLinear Equivariant Imaging (NLEI) with Spiral Subsampling.}
	\label{tab:sup_nl_ei_spiral_table_mae}

\end{table*}

%
\clearpage
\begin{table*}[t] 
	
	\centering
	
	\begin{tabular}{ | c | c | c | c | } 
		
		\hline \multicolumn{4}{| c |}{NLEI with Spiral Subsampling - Mean MAPE} \\ \hline
		
		$\alpha$ 				& T1 MAPE (\%) & T2 MAPE (\%) & PD MAPE (\%) \\ \hline
		 
		$\text{0}$				&	49.0090	&	1149.3624	&	16.0226	\\
		$\text{10}^{\text{-12}}$	&	4.7014	&	9.9862	&	14.0194	\\
		$\text{10}^{\text{-11}}$	&	4.4575	&	9.9056	&	11.4266	\\
		$\text{10}^{\text{-10}}$	&	4.2462	&	8.7898	&	5.7634	\\
		$\text{10}^{\text{-9}}$	&	4.2033	&	8.7624	&	6.2499	\\
		$\text{10}^{\text{-8}}$	&	4.2236	&	8.6272	&	4.8952	\\
		$\text{10}^{\text{-7}}$	&	4.3001	&	8.7374	&	6.1698	\\
		$\text{10}^{\text{-6}}$	&	4.2363	&	9.7763	&	5.6213	\\
		$\text{10}^{\text{-5}}$	&	4.1260	&	9.8378	&	8.1353	\\
		$\text{10}^{\text{-4}}$	&	4.2679	&	9.1630	&	8.0242	\\
		$\text{10}^{\text{-3}}$	&	4.1794	&	9.0589	&	5.9696	\\
		$\text{10}^{\text{-2}}$	&	4.5183	&	9.1437	&	7.7790	\\
		$\text{10}^{\text{-1}}$	&	5.1712	&	10.4649	&	8.3138	\\
		$\text{1}$				&	6.9669	&	14.0936	&	32.4467	\\
		$\text{10}^{\text{1}}$	&	12.2393	&	22.5962	&	62.7768	\\
		$\text{10}^{\text{2}}$	&	14.1843	&	23.3730	&	75.4141	\\
		$\text{10}^{\text{3}}$	&	13.8452	&	60.0660	&	36.9571	\\
		$\text{10}^{\text{4}}$	&	67.4086	&	29.5688	&	29.7432	\\
		$\text{10}^{\text{5}}$	&	32.6355	&	93.1951	&	95.7941	\\
		$\text{10}^{\text{6}}$	&	45.2356	&	40.6930	&	53.8938	\\
		\hline
	
	\end{tabular}
	
	\caption[...]{The average MAPE across 15 test slices for each QMap, T1, T2 and PD, against the Equivariant loss hyperparameter, $\alpha$, for NonLinear Equivariant Imaging (NLEI) with Spiral Subsampling.}
	\label{tab:sup_nl_ei_spiral_table_mape}

\end{table*}

%
\clearpage
\begin{table*}[t] 
	
	\centering
	
	\begin{tabular}{ | c | c | c | c | } 
		
		\hline \multicolumn{4}{| c |}{NLEI with Spiral Subsampling - Mean PSNR} \\ \hline
		
		$\alpha$ 				& T1 PSNR (dB) & T2 PSNR (dB) & PD PSNR (dB) \\ \hline
		 
		$\text{0}$				&	10.9272	&	3.5530	&	22.7114	\\
		$\text{10}^{\text{-12}}$	&	32.4374	&	35.7135	&	24.0572	\\
		$\text{10}^{\text{-11}}$	&	32.8425	&	36.2210	&	25.7408	\\
		$\text{10}^{\text{-10}}$	&	33.3695	&	36.8044	&	30.6028	\\
		$\text{10}^{\text{-9}}$	&	33.5523	&	36.7713	&	30.2546	\\
		$\text{10}^{\text{-8}}$	&	33.6557	&	36.7658	&	31.4679	\\
		$\text{10}^{\text{-7}}$	&	33.3289	&	36.5065	&	30.2682	\\
		$\text{10}^{\text{-6}}$	&	33.3965	&	35.1671	&	30.7866	\\
		$\text{10}^{\text{-5}}$	&	33.6677	&	35.8088	&	28.2647	\\
		$\text{10}^{\text{-4}}$	&	33.2579	&	36.3769	&	28.4831	\\
		$\text{10}^{\text{-3}}$	&	33.3009	&	36.7610	&	30.3547	\\
		$\text{10}^{\text{-2}}$	&	31.9512	&	36.4093	&	28.8938	\\
		$\text{10}^{\text{-1}}$	&	30.8338	&	35.8817	&	28.0999	\\
		$\text{1}$				&	28.4101	&	34.4717	&	16.9311	\\
		$\text{10}^{\text{1}}$		&	24.6209	&	31.9920	&	11.5577	\\
		$\text{10}^{\text{2}}$		&	23.8994	&	32.0548	&	9.4567	\\
		$\text{10}^{\text{3}}$		&	23.8594	&	30.4008	&	13.2923	\\
		$\text{10}^{\text{4}}$		&	13.1451	&	29.9750	&	16.8556	\\
		$\text{10}^{\text{5}}$		&	17.7444	&	25.6197	&	9.4317	\\
		$\text{10}^{\text{6}}$		&	15.8294	&	29.1467	&	13.4438	\\
		\hline
	
	\end{tabular}
	
	\caption[...]{The average PSNR across 15 test slices for each QMap, T1, T2 and PD, against the Equivariant loss hyperparameter, $\alpha$, for NonLinear Equivariant Imaging (NLEI) with Spiral Subsampling.}
	\label{tab:sup_nl_ei_spiral_table_psnr}

\end{table*}

%
\clearpage
\begin{table*}[t] 
	
	\centering
	
	\begin{tabular}{ | c | c | c | c | } 
		
		\hline \multicolumn{4}{| c |}{NLEI with Spiral Subsampling - Mean SSIM} \\ \hline
		
		$\alpha$ 				& T1 SSIM & T2 SSIM & PD SSIM \\ \hline
		 
		$\text{0}$				&	0.9170	&	0.9275	&	0.8114	\\
		$\text{10}^{\text{-12}}$	&	0.9287	&	0.9422	&	0.8589	\\
		$\text{10}^{\text{-11}}$	&	0.9347	&	0.9485	&	0.8672	\\
		$\text{10}^{\text{-10}}$	&	0.9402	&	0.9530	&	0.8891	\\
		$\text{10}^{\text{-9}}$	&	0.9414	&	0.9532	&	0.8901	\\
		$\text{10}^{\text{-8}}$	&	0.9425	&	0.9537	&	0.8913	\\
		$\text{10}^{\text{-7}}$	&	0.9394	&	0.9513	&	0.8873	\\
		$\text{10}^{\text{-6}}$	&	0.9400	&	0.9509	&	0.8872	\\
		$\text{10}^{\text{-5}}$	&	0.9433	&	0.9534	&	0.8846	\\
		$\text{10}^{\text{-4}}$	&	0.9389	&	0.9510	&	0.8798	\\
		$\text{10}^{\text{-3}}$	&	0.9393	&	0.9529	&	0.8862	\\
		$\text{10}^{\text{-2}}$	&	0.9248	&	0.9510	&	0.8803	\\
		$\text{10}^{\text{-1}}$	&	0.9044	&	0.9440	&	0.8655	\\
		$\text{1}$				&	0.8512	&	0.9196	&	0.8021	\\
		$\text{10}^{\text{1}}$		&	0.7764	&	0.8745	&	0.7328	\\
		$\text{10}^{\text{2}}$		&	0.7661	&	0.8787	&	0.7033	\\
		$\text{10}^{\text{3}}$		&	0.7671	&	0.8068	&	0.7761	\\
		$\text{10}^{\text{4}}$		&	0.4510	&	0.7478	&	0.6608	\\
		$\text{10}^{\text{5}}$		&	0.5970	&	0.2888	&	0.6467	\\
		$\text{10}^{\text{6}}$		&	0.4746	&	0.6062	&	0.6894	\\
		\hline
	
	\end{tabular}
	
	\caption[...]{The average SSIM across 15 test slices for each QMap, T1, T2 and PD, against the Equivariant loss hyperparameter, $\alpha$, for NonLinear Equivariant Imaging (NLEI) with Spiral Subsampling.}
	\label{tab:sup_nl_ei_spiral_table_ssim}

\end{table*}

%

%
\clearpage
\begin{table*}[t] 
	
	\centering
	
	\begin{tabular}{ | c | c | c | c | } 
		
		\hline \multicolumn{4}{| c |}{NLEI with EPI Subsampling - Mean MAE} \\ \hline
		
		$\alpha$ 				& T1 MAE (s) & T2 MAE (s) & PD (a.u) \\ \hline
		 
		$\text{0}$				&	0.0308	&	0.0176	&	0.0084	\\
		$\text{10}^{\text{-12}}$	&	0.0207	&	0.0139	&	0.0063	\\
		$\text{10}^{\text{-11}}$	&	0.0212	&	0.0155	&	0.0069	\\
		$\text{10}^{\text{-10}}$	&	0.0209	&	0.0159	&	0.0067	\\
		$\text{10}^{\text{-9}}$	&	0.0192	&	0.0140	&	0.0065	\\
		$\text{10}^{\text{-8}}$	&	0.0241	&	0.0162	&	0.0083	\\
		$\text{10}^{\text{-7}}$	&	0.0194	&	0.0139	&	0.0060	\\
		$\text{10}^{\text{-6}}$	&	0.0220	&	0.0145	&	0.0063	\\
		$\text{10}^{\text{-5}}$	&	0.0224	&	0.0158	&	0.0066	\\
		$\text{10}^{\text{-4}}$	&	0.0180	&	0.0139	&	0.0063	\\
		$\text{10}^{\text{-3}}$	&	0.0185	&	0.0135	&	0.0065	\\
		$\text{10}^{\text{-2}}$	&	0.0223	&	0.0167	&	0.0078	\\
		$\text{10}^{\text{-1}}$	&	0.0517	&	0.0203	&	0.0159	\\
		$\text{1}$				&	0.2465	&	0.0569	&	0.1832	\\
		$\text{10}^{\text{1}}$		&	0.3509	&	0.0539	&	0.0716	\\
		$\text{10}^{\text{2}}$		&	0.4396	&	0.0494	&	0.0588	\\
		$\text{10}^{\text{3}}$		&	0.9787	&	0.2618	&	0.0594	\\
		$\text{10}^{\text{4}}$		&	0.5131	&	0.0663	&	0.2968	\\
		$\text{10}^{\text{5}}$		&	0.5092	&	0.0505	&	0.2831	\\
		$\text{10}^{\text{6}}$		&	0.5667	&	0.0393	&	0.2615	\\
		\hline
	
	\end{tabular}
	
	\caption[...]{The average MAE across 15 test slices for each QMap, T1, T2 and PD, against the Equivariant loss hyperparameter, $\alpha$, for NonLinear Equivariant Imaging (NLEI) with EPI Subsampling.}
	\label{tab:sup_nl_ei_epi_table_mae}

\end{table*}

%
\clearpage
\begin{table*}[t] 
	
	\centering
	
	\begin{tabular}{ | c | c | c | c | } 
		
		\hline \multicolumn{4}{| c |}{NLEI with EPI Subsampling - Mean MAPE} \\ \hline
		
		$\alpha$ 				& T1 MAPE (\%) & T2 MAPE (\%) & MAPE (\%) \\ \hline
		 
		$\text{0}$				&	2.3088	&	8.3396	&	3.1016	\\
		$\text{10}^{\text{-12}}$	&	1.5280	&	6.6779	&	2.2962	\\
		$\text{10}^{\text{-11}}$	&	1.5429	&	7.3496	&	2.5243	\\
		$\text{10}^{\text{-10}}$	&	1.5104	&	7.2350	&	2.4924	\\
		$\text{10}^{\text{-9}}$	&	1.4179	&	7.0652	&	2.3705	\\
		$\text{10}^{\text{-8}}$	&	1.8458	&	8.0811	&	2.9816	\\
		$\text{10}^{\text{-7}}$	&	1.4208	&	6.8627	&	2.1973	\\
		$\text{10}^{\text{-6}}$	&	1.5034	&	7.1576	&	2.3248	\\
		$\text{10}^{\text{-5}}$	&	1.5572	&	7.8345	&	2.5453	\\
		$\text{10}^{\text{-4}}$	&	1.2951	&	6.7534	&	2.3250	\\
		$\text{10}^{\text{-3}}$	&	1.3202	&	7.1090	&	2.3663	\\
		$\text{10}^{\text{-2}}$	&	1.6814	&	8.9584	&	2.8076	\\
		$\text{10}^{\text{-1}}$	&	4.1571	&	14.1638	&	6.1689	\\
		$\text{1}$				&	15.9551	&	34.7221	&	72.1526	\\
		$\text{10}^{\text{1}}$		&	24.8077	&	32.4086	&	28.4549	\\
		$\text{10}^{\text{2}}$		&	32.8733	&	28.8645	&	22.3299	\\
		$\text{10}^{\text{3}}$		&	82.7953	&	234.8664	&	22.9091	\\
		$\text{10}^{\text{4}}$		&	39.7339	&	44.6247	&	118.8180	\\
		$\text{10}^{\text{5}}$		&	39.3725	&	29.1175	&	113.5092	\\
		$\text{10}^{\text{6}}$		&	44.7012	&	18.7333	&	105.3424	\\
		\hline
	
	\end{tabular}
	
	\caption[...]{The average MAPE across 15 test slices for each QMap, T1, T2 and PD, against the Equivariant loss hyperparameter, $\alpha$, for NonLinear Equivariant Imaging (NLEI) with EPI Subsampling.}
	\label{tab:sup_nl_ei_epi_table_mape}

\end{table*}

%
\clearpage
\begin{table*}[t] 
	
	\centering
	
	\begin{tabular}{ | c | c | c | c | } 
		
		\hline \multicolumn{4}{| c |}{NLEI with EPI Subsampling - Mean PSNR} \\ \hline
		
		$\alpha$ 				& T1 PSNR (dB) & T2 PSNR (dB) & PSNR (dB) \\ \hline
		 
		$\text{0}$				&	36.9056	&	35.4542	&	35.8991	\\
		$\text{10}^{\text{-12}}$	&	38.9810	&	37.1024	&	38.3585	\\
		$\text{10}^{\text{-11}}$	&	38.2721	&	36.3653	&	37.3931	\\
		$\text{10}^{\text{-10}}$	&	38.3038	&	35.9135	&	37.8005	\\
		$\text{10}^{\text{-9}}$	&	39.0430	&	37.0306	&	37.7414	\\
		$\text{10}^{\text{-8}}$	&	37.8366	&	36.2374	&	36.0738	\\
		$\text{10}^{\text{-7}}$	&	39.0094	&	37.0378	&	38.4272	\\
		$\text{10}^{\text{-6}}$	&	37.3860	&	36.6901	&	38.1582	\\
		$\text{10}^{\text{-5}}$	&	37.7602	&	36.0914	&	37.4961	\\
		$\text{10}^{\text{-4}}$	&	39.2725	&	37.0679	&	38.2944	\\
		$\text{10}^{\text{-3}}$	&	39.0171	&	37.5050	&	38.2829	\\
		$\text{10}^{\text{-2}}$	&	38.0982	&	35.9291	&	36.4459	\\
		$\text{10}^{\text{-1}}$	&	33.4653	&	35.5957	&	30.4087	\\
		$\text{1}$				&	21.2098	&	29.3584	&	10.6199	\\
		$\text{10}^{\text{1}}$		&	19.1448	&	29.7561	&	17.5979	\\
		$\text{10}^{\text{2}}$		&	17.6816	&	30.1725	&	18.9534	\\
		$\text{10}^{\text{3}}$		&	11.6431	&	19.5159	&	18.8848	\\
		$\text{10}^{\text{4}}$		&	16.6225	&	28.8932	&	6.7564	\\
		$\text{10}^{\text{5}}$		&	16.6768	&	30.0111	&	7.1636	\\
		$\text{10}^{\text{6}}$		&	15.9038	&	30.8007	&	8.0054	\\
		\hline
	
	\end{tabular}
	
	\caption[...]{The average PSNR across 15 test slices for each QMap, T1, T2 and PD, against the Equivariant loss hyperparameter, $\alpha$, for NonLinear Equivariant Imaging (NLEI) with EPI Subsampling.}
	\label{tab:sup_nl_ei_epi_table_psnr}

\end{table*}

%
\clearpage
\begin{table*}[t] 
	
	\centering
	
	\begin{tabular}{ | c | c | c | c | } 
		
		\hline \multicolumn{4}{| c |}{NLEI with EPI Subsampling - Mean SSIM} \\ \hline
		
		$\alpha$ 				& T1 SSIM & T2 SSIM & PD SSIM \\ \hline
		 
		$\text{0}$				&	0.9843	&	0.9478	&	0.9732	\\
		$\text{10}^{\text{-12}}$	&	0.9894	&	0.9646	&	0.9844	\\
		$\text{10}^{\text{-11}}$	&	0.9876	&	0.9573	&	0.9821	\\
		$\text{10}^{\text{-10}}$	&	0.9875	&	0.9544	&	0.9820	\\
		$\text{10}^{\text{-9}}$	&	0.9888	&	0.9636	&	0.9826	\\
		$\text{10}^{\text{-8}}$	&	0.9858	&	0.9552	&	0.9735	\\
		$\text{10}^{\text{-7}}$	&	0.9894	&	0.9640	&	0.9834	\\
		$\text{10}^{\text{-6}}$	&	0.9869	&	0.9614	&	0.9826	\\
		$\text{10}^{\text{-5}}$	&	0.9872	&	0.9555	&	0.9812	\\
		$\text{10}^{\text{-4}}$	&	0.9898	&	0.9642	&	0.9837	\\
		$\text{10}^{\text{-3}}$	&	0.9898	&	0.9660	&	0.9836	\\
		$\text{10}^{\text{-2}}$	&	0.9877	&	0.9540	&	0.9744	\\
		$\text{10}^{\text{-1}}$	&	0.9499	&	0.9443	&	0.9126	\\
		$\text{1}$				&	0.7553	&	0.6870	&	0.6761	\\
		$\text{10}^{\text{1}}$		&	0.6820	&	0.7155	&	0.6599	\\
		$\text{10}^{\text{2}}$		&	0.5933	&	0.7536	&	0.6510	\\
		$\text{10}^{\text{3}}$		&	0.4642	&	0.4123	&	0.6470	\\
		$\text{10}^{\text{4}}$		&	0.5184	&	0.5524	&	0.6202	\\
		$\text{10}^{\text{5}}$		&	0.5220	&	0.7527	&	0.6245	\\
		$\text{10}^{\text{6}}$		&	0.4779	&	0.8412	&	0.6315	\\
		\hline
	
	\end{tabular}
	
	\caption[...]{The average SSIM across 15 test slices for each QMap, T1, T2 and PD, against the Equivariant loss hyperparameter, $\alpha$, for NonLinear Equivariant Imaging (NLEI) with EPI Subsampling.}
	\label{tab:sup_nl_ei_epi_table_ssim}

\end{table*}

%

%
\clearpage
\begin{table*}[t] 
	
	\centering
	
	\begin{tabular}{ | c | c | c | c | } 
		
		\hline \multicolumn{4}{| c |}{EI with Spiral Subsampling - Mean MAE} \\ \hline
		
		$\alpha$ 				& T1 MAE (s) & T2 MAE (s) & PD MAE (a.u) \\ \hline
		 
		$\text{0}$				&	0.0538	&	0.0164	&	0.0147	\\
		$\text{10}^{\text{-12}}$	&	0.0521	&	0.0159	&	0.0166	\\
		$\text{10}^{\text{-11}}$	&	0.0535	&	0.0161	&	0.0162	\\
		$\text{10}^{\text{-10}}$	&	0.0545	&	0.0163	&	0.0163	\\
		$\text{10}^{\text{-9}}$	&	0.0523	&	0.0159	&	0.0174	\\
		$\text{10}^{\text{-8}}$	&	0.0563	&	0.0178	&	0.0229	\\
		$\text{10}^{\text{-7}}$	&	0.0540	&	0.0167	&	0.0177	\\
		$\text{10}^{\text{-6}}$	&	0.0533	&	0.0163	&	0.0153	\\
		$\text{10}^{\text{-5}}$	&	0.0506	&	0.0157	&	0.0135	\\
		$\text{10}^{\text{-4}}$	&	0.0525	&	0.0163	&	0.0171	\\
		$\text{10}^{\text{-3}}$	&	0.0518	&	0.0158	&	0.0146	\\
		$\text{10}^{\text{-2}}$	&	0.0530	&	0.0164	&	0.0153	\\
		$\text{10}^{\text{-1}}$	&	0.0543	&	0.0165	&	0.0223	\\
		$\text{1}$				&	0.0601	&	0.0192	&	0.0231	\\
		$\text{10}^{\text{1}}$		&	0.0734	&	0.0209	&	0.0499	\\
		$\text{10}^{\text{2}}$		&	0.1451	&	0.0315	&	0.1849	\\
		$\text{10}^{\text{3}}$		&	0.1767	&	0.0315	&	0.1737	\\
		$\text{10}^{\text{4}}$		&	0.1777	&	0.0310	&	0.1050	\\
		$\text{10}^{\text{5}}$		&	0.1798	&	0.0305	&	0.2523	\\
		$\text{10}^{\text{6}}$		&	0.1799	&	0.0305	&	0.2931	\\
		\hline
	
	\end{tabular}
	
	\caption[...]{The average MAE across 15 test slices for each QMap, T1, T2 and PD, against the Equivariant loss hyperparameter, $\alpha$, for linear Equivariant Imaging (EI) with Spiral Subsampling.}
	\label{tab:sup_ei_spiral_table_mae}

\end{table*}

%
\clearpage
\begin{table*}[t] 
	
	\centering
	
	\begin{tabular}{ | c | c | c | c | } 
		
		\hline \multicolumn{4}{| c |}{EI with Spiral Subsampling - Mean MAPE} \\ \hline
		
		$\alpha$ 				& T1 MAPE (\%) & T2 MAPE (\%) & PD MAPE (\%) \\ \hline
		 
		$\text{0}$				&	4.2072	&	8.7901	&	5.6714	\\
		$\text{10}^{\text{-12}}$	&	4.0780	&	8.6528	&	6.4161	\\
		$\text{10}^{\text{-11}}$	&	4.1792	&	8.8209	&	6.2615	\\
		$\text{10}^{\text{-10}}$	&	4.3127	&	9.1114	&	6.1888	\\
		$\text{10}^{\text{-9}}$	&	4.1027	&	8.5670	&	6.7515	\\
		$\text{10}^{\text{-8}}$	&	4.4946	&	9.7638	&	8.6750	\\
		$\text{10}^{\text{-7}}$	&	4.2808	&	8.9900	&	6.9081	\\
		$\text{10}^{\text{-6}}$	&	4.2826	&	8.7033	&	5.8529	\\
		$\text{10}^{\text{-5}}$	&	4.0897	&	8.4305	&	5.2789	\\
		$\text{10}^{\text{-4}}$	&	4.1112	&	8.4252	&	6.5966	\\
		$\text{10}^{\text{-3}}$	&	4.0683	&	8.6273	&	5.6712	\\
		$\text{10}^{\text{-2}}$	&	4.1309	&	8.7967	&	5.9373	\\
		$\text{10}^{\text{-1}}$	&	4.2045	&	8.9071	&	8.5592	\\
		$\text{1}$				&	4.7045	&	10.4998	&	8.8462	\\
		$\text{10}^{\text{1}}$		&	5.6234	&	10.2978	&	19.2170	\\
		$\text{10}^{\text{2}}$		&	11.1458	&	16.9386	&	72.8756	\\
		$\text{10}^{\text{3}}$		&	14.3074	&	15.6273	&	67.6573	\\
		$\text{10}^{\text{4}}$		&	15.0264	&	16.9562	&	41.6878	\\
		$\text{10}^{\text{5}}$		&	15.6406	&	14.4145	&	101.4867	\\
		$\text{10}^{\text{6}}$		&	15.6578	&	14.6547	&	117.4045	\\
		\hline
	
	\end{tabular}
	
	\caption[...]{The average MAPE across 15 test slices for each QMap, T1, T2 and PD, against the Equivariant loss hyperparameter, $\alpha$, for linear Equivariant Imaging (EI) with Spiral Subsampling.}
	\label{tab:sup_ei_spiral_table_mape}

\end{table*}

%
\clearpage
\begin{table*}[t] 
	
	\centering
	
	\begin{tabular}{ | c | c | c | c | c |} 
		
		\hline \multicolumn{5}{| c |}{EI with Spiral Subsampling - Mean PSNR} \\ \hline
		
		$\alpha$ 				& T1 PSNR (dB) & T2 PSNR (dB) & PD PSNR (dB) & TSMI PSNR (dB) \\ \hline
		 
		$\text{0}$				&	33.0338	&	36.1964	&	30.7113	&	18.8197	\\
		$\text{10}^{\text{-12}}$	&	33.3766	&	36.8253	&	30.0492	&	21.0106	\\
		$\text{10}^{\text{-11}}$	&	32.9535	&	36.3929	&	30.2479	&	21.4194	\\
		$\text{10}^{\text{-10}}$	&	32.7302	&	36.4768	&	30.0981	&	21.3871	\\
		$\text{10}^{\text{-9}}$	&	33.3861	&	36.5719	&	29.7736	&	21.5697	\\
		$\text{10}^{\text{-8}}$	&	32.3717	&	35.4812	&	27.2573	&	18.6427	\\
		$\text{10}^{\text{-7}}$	&	32.7760	&	36.0602	&	29.6220	&	21.4802	\\
		$\text{10}^{\text{-6}}$	&	33.2046	&	36.2564	&	30.5388	&	21.3567	\\
		$\text{10}^{\text{-5}}$	&	33.5820	&	36.7728	&	31.4401	&	21.3110	\\
		$\text{10}^{\text{-4}}$	&	33.2867	&	36.3890	&	29.8238	&	21.7094	\\
		$\text{10}^{\text{-3}}$	&	33.5632	&	36.7695	&	30.9022	&	21.9795	\\
		$\text{10}^{\text{-2}}$	&	33.1167	&	36.2874	&	30.5400	&	21.5254	\\
		$\text{10}^{\text{-1}}$	&	32.9669	&	36.4769	&	27.9505	&	21.8362	\\
		$\text{1}$				&	31.5910	&	33.8421	&	27.6364	&	19.2568	\\
		$\text{10}^{\text{1}}$		&	30.7331	&	34.4141	&	21.5620	&	21.8899	\\
		$\text{10}^{\text{2}}$		&	25.2856	&	31.7330	&	10.3843	&	17.8771	\\
		$\text{10}^{\text{3}}$		&	23.8945	&	31.6696	&	10.2516	&	17.3663	\\
		$\text{10}^{\text{4}}$		&	24.0135	&	31.8874	&	14.2755	&	19.1679	\\
		$\text{10}^{\text{5}}$		&	24.0452	&	31.7073	&	8.1859	&	20.0274	\\
		$\text{10}^{\text{6}}$		&	24.0428	&	31.7152	&	6.8634	&	19.9352	\\
		\hline
	
	\end{tabular}
	
	\caption[...]{The average PSNR across 15 test slices for each QMap, T1, T2 and PD, against the Equivariant loss hyperparameter, $\alpha$, for linear Equivariant Imaging (EI) with Spiral Subsampling.}
	\label{tab:sup_ei_spiral_table_psnr}

\end{table*}

%
\clearpage
\begin{table*}[t] 
	
	\centering
	
	\begin{tabular}{ | c | c | c | c | c |} 
		
		\hline \multicolumn{5}{| c |}{EI with Spiral Subsampling - Mean SSIM} \\ \hline
		
		$\alpha$ 				& T1 SSIM & T2 SSIM & PD SSIM & TSMI SSIM \\ \hline
		 
		$\text{0}$				&	0.9392	&	0.9490	&	0.8872	&	0.6596	\\
		$\text{10}^{\text{-12}}$	&	0.9432	&	0.9530	&	0.8927	&	0.6612	\\
		$\text{10}^{\text{-11}}$	&	0.9407	&	0.9515	&	0.8909	&	0.6605	\\
		$\text{10}^{\text{-10}}$	&	0.9378	&	0.9513	&	0.8884	&	0.6596	\\
		$\text{10}^{\text{-9}}$	&	0.9429	&	0.9520	&	0.8940	&	0.6619	\\
		$\text{10}^{\text{-8}}$	&	0.9364	&	0.9429	&	0.8783	&	0.6555	\\
		$\text{10}^{\text{-7}}$	&	0.9392	&	0.9486	&	0.8891	&	0.6622	\\
		$\text{10}^{\text{-6}}$	&	0.9412	&	0.9503	&	0.8922	&	0.6588	\\
		$\text{10}^{\text{-5}}$	&	0.9464	&	0.9530	&	0.8969	&	0.6581	\\
		$\text{10}^{\text{-4}}$	&	0.9416	&	0.9496	&	0.8912	&	0.6652	\\
		$\text{10}^{\text{-3}}$	&	0.9439	&	0.9534	&	0.8941	&	0.6680	\\
		$\text{10}^{\text{-2}}$	&	0.9407	&	0.9494	&	0.8928	&	0.6613	\\
		$\text{10}^{\text{-1}}$	&	0.9392	&	0.9502	&	0.8855	&	0.6676	\\
		$\text{1}$				&	0.9261	&	0.9358	&	0.8757	&	0.6718	\\
		$\text{10}^{\text{1}}$		&	0.8948	&	0.9235	&	0.8502	&	0.6732	\\
		$\text{10}^{\text{2}}$		&	0.7862	&	0.8817	&	0.7137	&	0.6492	\\
		$\text{10}^{\text{3}}$		&	0.7649	&	0.8809	&	0.7085	&	0.6392	\\
		$\text{10}^{\text{4}}$		&	0.7650	&	0.8860	&	0.7543	&	0.6373	\\
		$\text{10}^{\text{5}}$		&	0.7639	&	0.8857	&	0.6392	&	0.6186	\\
		$\text{10}^{\text{6}}$		&	0.7638	&	0.8855	&	0.6216	&	0.6151	\\
		\hline
	
	\end{tabular}
	
	\caption[...]{The average SSIM across 15 test slices for each QMap, T1, T2 and PD, against the Equivariant loss hyperparameter, $\alpha$, for linear Equivariant Imaging (EI) with Spiral Subsampling.}
	\label{tab:sup_ei_spiral_table_ssim}

\end{table*}

%

%
\clearpage
\begin{table*}[t] 
	
	\centering
	
	\begin{tabular}{ | c | c | c | c | } 
		
		\hline \multicolumn{4}{| c |}{EI with EPI Subsampling - Mean MAE} \\ \hline
		
		$\alpha$ 				& T1 MAE (s) & T2 MAE (s) & PD MAE (a.u) \\ \hline
		 
		$\text{0}$				&	0.3375	&	0.1650	&	0.0774	\\
		$\text{10}^{\text{-12}}$	&	0.3554	&	0.0831	&	0.0756	\\
		$\text{10}^{\text{-11}}$	&	0.2395	&	0.0967	&	0.0718	\\
		$\text{10}^{\text{-10}}$	&	0.2732	&	0.0882	&	0.0917	\\
		$\text{10}^{\text{-9}}$	&	0.2022	&	0.0674	&	0.0886	\\
		$\text{10}^{\text{-8}}$	&	0.1943	&	0.2332	&	0.0731	\\
		$\text{10}^{\text{-7}}$	&	0.1964	&	0.1814	&	0.0381	\\
		$\text{10}^{\text{-6}}$	&	0.2449	&	0.2249	&	0.0279	\\
		$\text{10}^{\text{-5}}$	&	0.1675	&	0.2048	&	0.0348	\\
		$\text{10}^{\text{-4}}$	&	0.2449	&	0.0815	&	0.0498	\\
		$\text{10}^{\text{-3}}$	&	0.2806	&	0.0931	&	0.0457	\\
		$\text{10}^{\text{-2}}$	&	0.1096	&	0.0451	&	0.0268	\\
		$\text{10}^{\text{-1}}$	&	0.0844	&	0.2055	&	0.0142	\\
		$\text{1}$				&	0.0950	&	0.0478	&	0.0365	\\
		$\text{10}^{\text{1}}$		&	0.1194	&	0.1795	&	0.1412	\\
		$\text{10}^{\text{2}}$		&	0.4487	&	0.3139	&	0.1347	\\
		$\text{10}^{\text{3}}$		&	1.4173	&	0.0372	&	0.1722	\\
		$\text{10}^{\text{4}}$		&	0.1861	&	0.4790	&	0.2943	\\
		$\text{10}^{\text{5}}$		&	0.1949	&	0.5075	&	0.2959	\\
		$\text{10}^{\text{6}}$		&	0.3137	&	0.0589	&	0.2968	\\
		\hline
	
	\end{tabular}
	
	\caption[...]{The average MAE across 15 test slices for each QMap, T1, T2 and PD, against the Equivariant loss hyperparameter, $\alpha$, for linear Equivariant Imaging (EI) with EPI Subsampling.}
	\label{tab:sup_ei_epi_table_mae}

\end{table*}

%
\clearpage
\begin{table*}[t] 
	
	\centering
	
	\begin{tabular}{ | c | c | c | c | } 
		
		\hline \multicolumn{4}{| c |}{EI with EPI Subsampling - Mean MAPE} \\ \hline
		
		$\alpha$ 				& T1 MAPE (\%) & T2 MAPE (\%) & PD MAPE (\%) \\ \hline
		 
		$\text{0}$				&	26.3318	&	135.5899	&	27.7157	\\
		$\text{10}^{\text{-12}}$	&	26.1461	&	74.1114	&	26.7611	\\
		$\text{10}^{\text{-11}}$	&	18.2042	&	99.2322	&	27.3531	\\
		$\text{10}^{\text{-10}}$	&	19.0999	&	67.3423	&	36.7881	\\
		$\text{10}^{\text{-9}}$	&	14.3593	&	50.3731	&	31.6792	\\
		$\text{10}^{\text{-8}}$	&	15.3900	&	177.1664	&	27.2429	\\
		$\text{10}^{\text{-7}}$	&	13.2067	&	173.7378	&	13.1139	\\
		$\text{10}^{\text{-6}}$	&	17.3777	&	183.5313	&	10.1765	\\
		$\text{10}^{\text{-5}}$	&	11.5588	&	148.5074	&	13.4065	\\
		$\text{10}^{\text{-4}}$	&	16.2921	&	69.1706	&	19.4256	\\
		$\text{10}^{\text{-3}}$	&	20.7409	&	70.8022	&	17.7070	\\
		$\text{10}^{\text{-2}}$	&	7.4711	&	33.5549	&	10.6282	\\
		$\text{10}^{\text{-1}}$	&	6.0060	&	191.1255	&	5.4268	\\
		$\text{1}$				&	8.7539	&	37.2691	&	14.7984	\\
		$\text{10}^{\text{1}}$		&	9.6941	&	142.8419	&	56.2033	\\
		$\text{10}^{\text{2}}$		&	37.9268	&	309.2443	&	53.9230	\\
		$\text{10}^{\text{3}}$		&	141.3228	&	22.9724	&	71.1500	\\
		$\text{10}^{\text{4}}$		&	16.9006	&	485.8691	&	117.8429	\\
		$\text{10}^{\text{5}}$		&	18.3943	&	513.8505	&	118.4888	\\
		$\text{10}^{\text{6}}$		&	33.1389	&	37.1800	&	118.8190	\\
		\hline
	
	\end{tabular}
	
	\caption[...]{The average MAPE across 15 test slices for each QMap, T1, T2 and PD, against the Equivariant loss hyperparameter, $\alpha$, for linear Equivariant Imaging (EI) with EPI Subsampling.}
	\label{tab:sup_ei_epi_table_mape}

\end{table*}

%
\clearpage
\begin{table*}[t] 
	
	\centering
	
	\begin{tabular}{ | c | c | c | c | c |} 
		
		\hline \multicolumn{5}{| c |}{EI with EPI Subsampling - Mean PSNR} \\ \hline
		
		$\alpha$ 				& T1 PSNR (dB) & T2 PSNR (dB) & PD PSNR (dB) & TSMI PSNR (dB) \\ \hline
		 
		$\text{0}$				&	18.2728	&	20.7138	&	17.8021	&	0.3392	\\
		$\text{10}^{\text{-12}}$	&	17.6452	&	24.0347	&	18.0257	&	0.4167	\\
		$\text{10}^{\text{-11}}$	&	18.9077	&	27.0630	&	19.4534	&	-5.1060	\\
		$\text{10}^{\text{-10}}$	&	19.5507	&	23.1897	&	15.6966	&	0.6164	\\
		$\text{10}^{\text{-9}}$	&	19.6810	&	22.6231	&	16.7673	&	1.6752	\\
		$\text{10}^{\text{-8}}$	&	21.6021	&	17.4648	&	18.9366	&	-2.7664	\\
		$\text{10}^{\text{-7}}$	&	20.4500	&	20.7188	&	23.2824	&	2.5408	\\
		$\text{10}^{\text{-6}}$	&	20.5135	&	18.2230	&	25.0303	&	3.9077	\\
		$\text{10}^{\text{-5}}$	&	23.3167	&	18.1354	&	22.9783	&	3.3348	\\
		$\text{10}^{\text{-4}}$	&	19.7696	&	26.0009	&	20.6255	&	2.8656	\\
		$\text{10}^{\text{-3}}$	&	19.8411	&	23.8267	&	21.4395	&	0.6290	\\
		$\text{10}^{\text{-2}}$	&	23.3885	&	31.1333	&	25.9406	&	3.2324	\\
		$\text{10}^{\text{-1}}$	&	28.6166	&	20.2872	&	30.3307	&	0.5505	\\
		$\text{1}$				&	25.4382	&	24.6760	&	23.7842	&	1.5944	\\
		$\text{10}^{\text{1}}$		&	25.2234	&	19.1101	&	12.6983	&	1.4202	\\
		$\text{10}^{\text{2}}$		&	13.9856	&	15.7975	&	12.8303	&	2.6748	\\
		$\text{10}^{\text{3}}$		&	7.5946	&	31.4068	&	11.0399	&	3.1600	\\
		$\text{10}^{\text{4}}$		&	24.0091	&	14.5836	&	6.8287	&	3.3823	\\
		$\text{10}^{\text{5}}$		&	23.8720	&	14.0852	&	6.7803	&	-0.4833	\\
		$\text{10}^{\text{6}}$		&	20.8934	&	29.4065	&	6.7557	&	-10.1756	\\
		\hline
	
	\end{tabular}
	
	\caption[...]{The average PSNR across 15 test slices for each QMap, T1, T2 and PD, against the Equivariant loss hyperparameter, $\alpha$, for linear Equivariant Imaging (EI) with EPI Subsampling.}
	\label{tab:sup_ei_epi_table_psnr}

\end{table*}

%
\clearpage
\begin{table*}[t] 
	
	\centering
	
	\begin{tabular}{ | c | c | c | c | c |} 
		
		\hline \multicolumn{5}{| c |}{EI with EPI Subsampling - Mean SSIM} \\ \hline
		
		$\alpha$ 				& T1 SSIM & T2 SSIM & PD SSIM & TSMI SSIM \\ \hline
		 
		$\text{0}$				&	0.7302	&	0.6198	&	0.7043	&	0.5373	\\
		$\text{10}^{\text{-12}}$	&	0.7340	&	0.6823	&	0.6956	&	0.5318	\\
		$\text{10}^{\text{-11}}$	&	0.8383	&	0.7493	&	0.6976	&	0.5364	\\
		$\text{10}^{\text{-10}}$	&	0.8076	&	0.6608	&	0.6787	&	0.5368	\\
		$\text{10}^{\text{-9}}$	&	0.8574	&	0.8093	&	0.6377	&	0.5501	\\
		$\text{10}^{\text{-8}}$	&	0.8859	&	0.6241	&	0.6601	&	0.5373	\\
		$\text{10}^{\text{-7}}$	&	0.8545	&	0.6615	&	0.8607	&	0.5437	\\
		$\text{10}^{\text{-6}}$	&	0.8368	&	0.6112	&	0.8329	&	0.5493	\\
		$\text{10}^{\text{-5}}$	&	0.9020	&	0.6766	&	0.8084	&	0.5519	\\
		$\text{10}^{\text{-4}}$	&	0.8425	&	0.7162	&	0.7690	&	0.5457	\\
		$\text{10}^{\text{-3}}$	&	0.8066	&	0.7049	&	0.7367	&	0.5424	\\
		$\text{10}^{\text{-2}}$	&	0.9230	&	0.8376	&	0.9050	&	0.5542	\\
		$\text{10}^{\text{-1}}$	&	0.9521	&	0.6520	&	0.9507	&	0.5421	\\
		$\text{1}$				&	0.9194	&	0.8535	&	0.8451	&	0.5499	\\
		$\text{10}^{\text{1}}$		&	0.8714	&	0.6989	&	0.6805	&	0.5530	\\
		$\text{10}^{\text{2}}$		&	0.6772	&	0.5798	&	0.7001	&	0.5542	\\
		$\text{10}^{\text{3}}$		&	0.5863	&	0.8663	&	0.6865	&	0.5486	\\
		$\text{10}^{\text{4}}$		&	0.7612	&	0.5542	&	0.6212	&	0.5634	\\
		$\text{10}^{\text{5}}$		&	0.7580	&	0.5501	&	0.6205	&	0.5453	\\
		$\text{10}^{\text{6}}$		&	0.7258	&	0.6540	&	0.6202	&	0.5075	\\
		\hline
	
	\end{tabular}
	
	\caption[...]{The average SSIM across 15 test slices for each QMap, T1, T2 and PD, against the Equivariant loss hyperparameter, $\alpha$, for linear Equivariant Imaging (EI) with EPI Subsampling.}
	\label{tab:sup_ei_epi_table_ssim}

\end{table*}

%

\clearpage
%
\begin{table}[t] 
	
	\centering
	
	\begin{tabular}{ | c | c | c | c | c | } 
		
		\hline \multicolumn{5}{| c |}{NLEI with Spiral Subsampling - Mean QMaps} \\ \hline
		
		$\alpha$ 				& MAE (s) & MAPE (\%) & PSNR (dB) & SSIM \\ \hline
		 
		$\text{0}$				&	0.4328	&	404.7980	&	12.3972	&	0.8853	\\
		$\text{10}^{\text{-12}}$	&	0.0386	&	9.5690	&	30.7360	&	0.9099	\\
		$\text{10}^{\text{-11}}$	&	0.0350	&	8.5966	&	31.6014	&	0.9168	\\
		$\text{10}^{\text{-10}}$	&	0.0287	&	6.2665	&	33.5922	&	0.9274	\\
		$\text{10}^{\text{-9}}$	&	0.0286	&	6.4052	&	33.5261	&	0.9282	\\
		$\text{\bf{10}}^{\text{\bf{-8}}}$	&	\bf{0.0274}	&	\bf{5.9153}	&	\bf{33.9631}	&	\bf{0.9292}	\\
		$\text{10}^{\text{-7}}$	&	0.0290	&	6.4024	&	33.3679	&	0.9260	\\
		$\text{10}^{\text{-6}}$	&	0.0287	&	6.5446	&	33.1167	&	0.9260	\\
		$\text{10}^{\text{-5}}$	&	0.0303	&	7.3664	&	32.5804	&	0.9271	\\
		$\text{10}^{\text{-4}}$	&	0.0309	&	7.1517	&	32.7060	&	0.9233	\\
		$\text{10}^{\text{-3}}$	&	0.0288	&	6.4026	&	33.4722	&	0.9261	\\
		$\text{10}^{\text{-2}}$	&	0.0327	&	7.1470	&	32.4181	&	0.9187	\\
		$\text{10}^{\text{-1}}$	&	0.0369	&	7.9833	&	31.6051	&	0.9046	\\
		$\text{1}$				&	0.0681	&	17.8357	&	26.6043	&	0.8576	\\
		$\text{10}^{\text{1}}$	&	0.1173	&	32.5374	&	22.7236	&	0.7946	\\
		$\text{10}^{\text{2}}$	&	0.1343	&	37.6572	&	21.8036	&	0.7827	\\
		$\text{10}^{\text{3}}$	&	0.1117	&	36.9561	&	22.5175	&	0.7833	\\
		$\text{10}^{\text{4}}$	&	0.3141	&	42.2402	&	19.9919	&	0.6199	\\
		$\text{10}^{\text{5}}$	&	0.2620	&	73.8749	&	17.5986	&	0.5108	\\
		$\text{10}^{\text{6}}$	&	0.2546	&	46.6075	&	19.4733	&	0.5901	\\
		\hline
	
	\end{tabular}
	
	\caption[...]{The average result using all QMaps, T1, T2 and PD, (for 15 test slices) with respect to a particular metric against varying the Equivariant loss hyperparameter, $\alpha$, for NonLinear Equivariant Imaging (NLEI) with Spiral Subsampling.}
	\label{tab:sup_nl_ei_spiral_table_avg_qmaps}

\end{table}

%
\begin{table}[h] 
	
	\centering
	
	\begin{tabular}{ | c | c | c | c | c | } 
		
		\hline \multicolumn{5}{| c |}{NLEI with EPI Subsampling - Mean QMaps} \\ \hline
		
		$\alpha$ 				& MAE (s) & MAPE (\%) & PSNR (dB) & SSIM \\ \hline
		 
		$\text{0}$				&	0.0189	&	4.5833	&	36.0863	&	0.9685	\\
		$\text{10}^{\text{-12}}$	&	0.0136	&	3.5007	&	38.1473	&	0.9795	\\
		$\text{10}^{\text{-11}}$	&	0.0145	&	3.8056	&	37.3435	&	0.9757	\\
		$\text{10}^{\text{-10}}$	&	0.0145	&	3.7459	&	37.3393	&	0.9746	\\
		$\text{10}^{\text{-9}}$	&	0.0133	&	3.6179	&	37.9383	&	0.9783	\\
		$\text{10}^{\text{-8}}$	&	0.0162	&	4.3028	&	36.7159	&	0.9715	\\
		$\text{10}^{\text{-7}}$	&	0.0131	&	3.4936	&	38.1582	&	0.9789	\\
		$\text{10}^{\text{-6}}$	&	0.0143	&	3.6619	&	37.4114	&	0.9769	\\
		$\text{10}^{\text{-5}}$	&	0.0149	&	3.9790	&	37.1159	&	0.9746	\\
		$\text{\bf{10}}^{\text{\bf{-4}}}$	&	\bf{0.0127}	&	\bf{3.4579}	&	38.2116	&	0.9792	\\
		$\text{10}^{\text{-3}}$	&	0.0128	&	3.5985	&	\bf{38.2683}	&	\bf{0.9798}	\\
		$\text{10}^{\text{-2}}$	&	0.0156	&	4.4824	&	36.8244	&	0.9720	\\
		$\text{10}^{\text{-1}}$	&	0.0293	&	8.1633	&	33.1566	&	0.9356	\\
		$\text{1}$				&	0.1622	&	40.9433	&	20.3960	&	0.7061	\\
		$\text{10}^{\text{1}}$		&	0.1588	&	28.5571	&	22.1663	&	0.6858	\\
		$\text{10}^{\text{2}}$		&	0.1826	&	28.0226	&	22.2691	&	0.6660	\\
		$\text{10}^{\text{3}}$		&	0.4333	&	113.5236	&	16.6813	&	0.5078	\\
		$\text{10}^{\text{4}}$		&	0.2921	&	67.7255	&	17.4240	&	0.5637	\\
		$\text{10}^{\text{5}}$		&	0.2809	&	60.6664	&	17.9505	&	0.6331	\\
		$\text{10}^{\text{6}}$		&	0.2892	&	56.2590	&	18.2367	&	0.6502	\\
		\hline
	
	\end{tabular}
	
	\caption[...]{The average result using all QMaps, T1, T2 and PD, (for 15 test slices) with respect to a particular metric against varying the Equivariant loss hyperparameter, $\alpha$, for NonLinear Equivariant Imaging (NLEI) with EPI Subsampling.}
	\label{tab:sup_nl_ei_epi_table_avg_qmaps}

\end{table}

%

%
\begin{table}[h] 
	
	\centering
	
	\begin{tabular}{ | c | c | c | c | c | } 
		
		\hline \multicolumn{5}{| c |}{EI with Spiral Subsampling - Mean QMaps} \\ \hline
		
		$\alpha$ 				& MAE (s) & MAPE (\%) & PSNR (dB) & SSIM \\ \hline
		 
		$\text{0}$				&	0.0283	&	6.2229	&	33.3138	&	0.9252	\\
		$\text{10}^{\text{-12}}$	&	0.0282	&	6.3823	&	33.4170	&	0.9296	\\
		$\text{10}^{\text{-11}}$	&	0.0286	&	6.4206	&	33.1981	&	0.9277	\\
		$\text{10}^{\text{-10}}$	&	0.0290	&	6.5376	&	33.1017	&	0.9258	\\
		$\text{10}^{\text{-9}}$	&	0.0285	&	6.4737	&	33.2439	&	0.9296	\\
		$\text{10}^{\text{-8}}$	&	0.0323	&	7.6445	&	31.7034	&	0.9192	\\
		$\text{10}^{\text{-7}}$	&	0.0295	&	6.7263	&	32.8194	&	0.9257	\\
		$\text{10}^{\text{-6}}$	&	0.0283	&	6.2796	&	33.3333	&	0.9279	\\
		$\text{\bf{10}}^{\text{\bf{-5}}}$	&	\bf{0.0266}	&	\bf{5.9331}	&	\bf{33.9316}	&	\bf{0.9321}	\\
		$\text{10}^{\text{-4}}$	&	0.0286	&	6.3777	&	33.1665	&	0.9275	\\
		$\text{10}^{\text{-3}}$	&	0.0274	&	6.1223	&	33.7450	&	0.9305	\\
		$\text{10}^{\text{-2}}$	&	0.0282	&	6.2883	&	33.3147	&	0.9276	\\
		$\text{10}^{\text{-1}}$	&	0.0310	&	7.2236	&	32.4648	&	0.9250	\\
		$\text{1}$				&	0.0341	&	8.0168	&	31.0232	&	0.9125	\\
		$\text{10}^{\text{1}}$		&	0.0481	&	11.7127	&	28.9031	&	0.8895	\\
		$\text{10}^{\text{2}}$		&	0.1205	&	33.6533	&	22.4676	&	0.7939	\\
		$\text{10}^{\text{3}}$		&	0.1273	&	32.5307	&	21.9385	&	0.7848	\\
		$\text{10}^{\text{4}}$		&	0.1046	&	24.5568	&	23.3921	&	0.8018	\\
		$\text{10}^{\text{5}}$		&	0.1542	&	43.8473	&	21.3128	&	0.7629	\\
		$\text{10}^{\text{6}}$		&	0.1679	&	49.2390	&	20.8738	&	0.7570	\\
		\hline
	
	\end{tabular}
	
	\caption[...]{The average result using all QMaps, T1, T2 and PD, (for 15 test slices) with respect to a particular metric against varying the Equivariant loss hyperparameter, $\alpha$, for linear Equivariant Imaging (EI) with Spiral Subsampling.}
	\label{tab:sup_ei_spiral_table_avg_qmaps}

\end{table}

%
\begin{table}[h] 
	
	\centering
	
	\begin{tabular}{ | c | c | c | c | c | } 
		
		\hline \multicolumn{5}{| c |}{EI with EPI SubSampling - Mean QMaps} \\ \hline
		
		$\alpha$ 				& MAE (s) & MAPE (\%) & PSNR (dB) & SSIM \\ \hline
		 
		$\text{0}$				&	0.1933	&	63.2125	&	18.9296	&	0.1410	\\
		$\text{10}^{\text{-12}}$	&	0.1713	&	42.3396	&	19.9019	&	0.1529	\\
		$\text{10}^{\text{-11}}$	&	0.1360	&	48.2632	&	21.8081	&	0.2405	\\
		$\text{10}^{\text{-10}}$	&	0.1510	&	41.0768	&	19.4790	&	0.3307	\\
		$\text{10}^{\text{-9}}$	&	0.1194	&	32.1372	&	19.6905	&	0.3151	\\
		$\text{10}^{\text{-8}}$	&	0.1669	&	73.2664	&	19.3345	&	0.3099	\\
		$\text{10}^{\text{-7}}$	&	0.1386	&	66.6862	&	21.4837	&	0.3529	\\
		$\text{10}^{\text{-6}}$	&	0.1659	&	70.3618	&	21.2556	&	0.3636	\\
		$\text{10}^{\text{-5}}$	&	0.1357	&	57.8243	&	21.4768	&	0.3587	\\
		$\text{10}^{\text{-4}}$	&	0.1254	&	34.9628	&	22.1320	&	0.3412	\\
		$\text{10}^{\text{-3}}$	&	0.1398	&	36.4167	&	21.7024	&	0.3338	\\
		$\text{\bf{10}}^{\text{\bf{-2}}}$	&	0.0605	&	\bf{17.2181}	&	\bf{26.8208}	&	\bf{0.4651}	\\
		$\text{10}^{\text{-1}}$	&	0.1014	&	67.5194	&	26.4115	&	0.2606	\\
		$\text{1}$				&	\bf{0.0598}	&	20.2738	&	24.6328	&	0.2696	\\
		$\text{10}^{\text{1}}$	&	0.1467	&	69.5798	&	19.0106	&	0.1639	\\
		$\text{10}^{\text{2}}$	&	0.2991	&	133.6980	&	14.2045	&	0.0843	\\
		$\text{10}^{\text{3}}$	&	0.5422	&	78.4817	&	16.6804	&	0.1967	\\
		$\text{10}^{\text{4}}$	&	0.3198	&	206.8709	&	15.1405	&	0.0963	\\
		$\text{10}^{\text{5}}$	&	0.3328	&	216.9112	&	14.9125	&	0.0946	\\
		$\text{10}^{\text{6}}$	&	0.2231	&	63.0460	&	19.0185	&	0.2835	\\
		\hline
	
	\end{tabular}
	
	\caption[...]{The average result using all QMaps, T1, T2 and PD, (for 15 test slices) with respect to a particular metric against varying the Equivariant loss hyperparameter, $\alpha$, for linear Equivariant Imaging (EI) with EPI Subsampling.}
	\label{tab:sup_ei_epi_table_avg_qmaps}

\end{table}

%